Masters of Science in Computer Science and Engineering

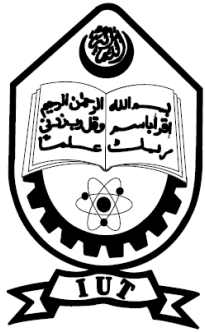

**Enhancing Software Development Process (ESDP) using Data Mining Integrated Environment**

by

Ziaur Rahman

Systems and Software Lab (SSL)

Supervisor

Md. Kamrul Hasan, PhD

Associate Professor, Department of CSE, IUT

Department of Computer Science and Engineering (CSE)

Islamic University of Technology (IUT)

October, 2015

# Form- A

The thesis titled "Enhancing Software Development Process (ESDP) using Data Mining Integrated Environment" Submitted by Ziaur Rahman St. No. 124606 of Academic Year 2014-2015 has been found satisfactory and is accepted as partial fulfillment of the requirement for the Degree of Masters of Science in Computer Science and Engineering on October 23, 2015.

## Board of Examiners

1. ……………………………………………………..
   Dr. Md. Kamrul Hasan, Supervisor            Chairman
   Associate Professor,
   Dept. of Computer Science and Engineering, IUT

2. ……………………………………………………..
   Prof. Dr. M. A. Mottalib                    Member
   Head,                                       (Ex-Officio)
   Dept. of Computer Science and Engineering, IUT

3. ……………………………………………………..
   Dr. Abu Raihan Mostofa Kamal                Member
   Associate Professor,
   Dept. of Computer Science and Engineering, IUT

4. ……………………………………………………..
   Dr. Md. Hasanul Kabir                       Member
   Associate Professor,
   Dept. of Computer Science and Engineering, IUT

5. ……………………………………………………..
   Prof. Dr. M. Kaykobad                       External Member
   Dean, Faculty of Electrical & Electronic Engineering,
   Bangladesh University of Engineering and Technology

# Form- B

## Declaration of Candidate

It is hereby declared that this thesis report or any part of it has not been submitted elsewhere for the award of any Degree or Diploma.

| | |
|---|---|
| Signature of the Supervisor | Signature of the Candidate |
| ………………………………… | ………………………………… |
| Md. Kamrul Hasan, PhD | Ziaur Rahman |
| Associate Professor | Systems and Software Lab (SSL) |
| Department of Computer Science and Engineering | Student No : 124606 |
| Islamic University of Technology (IUT), | Session     :  2011-2012 |
| Gazipur - 1704 | Date: October 23, 2015 |

# Dedication

*"This dissertation is dedicated to my parents and teachers for all their continuous support, love and inspiration"*

# Abstract

Nowadays, it has become a basic need to reuse existing Application Programming Interface (API), Class Libraries and frameworks for rapid software development. Software developers often reuse this by calling the respective APIs or libraries. But in doing so, developers usually encounter different problems in searching for appropriate code snippets. In most cases API and Libraries are complex and not well structured or well documented. Online search engine consumes time in searching, yet match is not that relevant and representation is not good. To get a suggestion according to the query we can find that snippet online using search engines or code search engines. In some cases database dependent searching and remote web server based mined repository searching bring problem to the developers. Finding an API recommendation on code search engine often deal with extra-large files that eventually slow down software development process. We have searched for a solution throughout our work and tried to bring a better outcome. As an alternative action we have implemented a system what we call "Enhancing Software Development Process (ESDP)" tool that is able to provide an efficient and working integrated environment to the developers with a better abstraction and representation of the search results and programmer's need to be derived from the source codes. We also have built and applied an XML based enriched repository to get recommendation from the mined repository in the client side without interacting with the internet dependent server to save complications and times. We provide the most relevant code skeletons or mapping to programmers/developers using graph based representation. We have evaluated that ESDP boosts up software development process enough particularly by reducing the required time in the coding phase. By giving a number of queries for an API, ESDP gathers more relevant source snippets through data mining. We have evaluated the efficiency of ESDP tool using a set of various queries and compared with the other existing tools. The results show that in different experiments ESDP consumes quite less time than some updated approaches to find the code snippet solution.

# Acknowledgments

I would like to thank and express my deepest appreciation and sincere gratitude to my supervisor Md. Kamrul Hasan, PhD, who inspired me to take pride in my research; his enthusiasm for research efforts surely will have a significant effect on my future research. His advice throughout this process kept me focused on the right direction. It is a great experience of my life time research. I am really thankful to him for sharing his ideas and necessary insights. I am highly grateful to Prof. Dr. M.A Mottalib, Head, Department of Computer Science and Engineering (CSE), IUT, for his encouragements and unique cooperation. I would like to especially thank to Dr. Hasanul Kabir and Hasan Mahmud for their kind appreciation. Lastly but not least, I would acknowledge and recall all of the teachers of IUT and my dear colleagues. They all supported me really a lot with their reviews, comments and suggestions during the research work.

Ziaur Rahman
*October 2015*

# Preface

This master's thesis is outlined based on Enhancing Software Development Process (ESDP) using data mining integrated environment. This research was carried out in the Systems and Software Lab (SSL) under the Department of Computer Science and Engineering (CSE) of Islamic University of Technology (IUT), OIC, Dhaka. This thesis includes five chapters which are briefed as follows:

**Chapter-1:** Chapter 1 provides a detailed discussion on the importance of the work that has been done and why the current topic is selected for Master's Thesis along with an informative introduction.

**Chapter-2:** Chapter 2 discusses about the background and related works to understand the ESDP Concepts. This also discusses the flaws of the existing approaches.

**Chapter-3:** Chapter 3 discusses about the proposed idea and the ESDP framework in details. It describes the Idea with the respective of four heuristics.

**Chapter-4:** Chapter 4 describes about the experimental evaluations. It shows the precision recall, Quantitative and Empirical evaluation.

**Chapter-5:** Chapter 5 provides conclusions to the discussions based on the ESDP concept and proves ideas for the future scope.

**Appendix**: Appendix covers information about some of the relevant technical experiments that has been done during research work.

# Contents

# List of Figures

# List of Tables

# List of Abbreviations

| | |
|---|---|
| **ESDP** | *Enhancing Software Development Process* |
| **MSR** | *Mining Software Repository* |
| **SR** | *Software Repository* |
| **DMIE** | *Data Mining Integrated Environment* |
| **MAPO** | *Mining API Usages from Open Source Repositories* |
| **MAC** | *Mining API Code for Code Reuse* |
| **API** | *Application Programming Interface* |
| **XML** | *Extensible Markup Language* |
| **CVS** | *Version Control System* |
| **BTS** | *Bug Tracking System* |
| **PD** | *Pattern Database* |
| **TD** | *Transaction Database* |
| **GUI** | *Graphical User Interface* |
| **USG Pattern** | *Usage Pattern* |
| **RAM** | *Random Access Memory* |
| **MGEF** | *My Eclipse Graphical Editing Framework* |
| **FP** | *Frequent Pattern* |
| **PD** | *Package Declaration* |
| **ID** | *Import Declaration* |
| **TD** | *Type Declaration* |
| **FD** | *Field Declaration* |
| **CI** | *Class Instance Creation* |
| **MD** | *Method Declaration* |
| **MI** | *Method Invocation* |
| **VD** | *Variable Declaration* |
| **ACD** | *Anonymous Class Declaration* |
| **AA** | *Array Access* |
| **AC** | *Array Creation* |
| **CTI** | *Constructor Invocation* |
| **FA** | *Field Access* |
| **SCI** | *Super Constructor Invocation* |
| **RT** | *Return Statement* |
| **SC** | *Super Class Inheritance* |
| **PrefixSpan** | *Prefix-Projected Sequential Pattern Mining* |
| **UQ** | *User Queries* |
| **ROC** | *Receiving Operating Characteristics* |
| **AUC** | *Area Under Curve* |
| **ACM** | *Association for Computing Machinery* |
| **SIGSOFT** | *Special Interest Group on Software Engineering* |
| **SIGKDD** | *Special Interest Group on Knowledge Discovery* |
| **IEEE** | *Institute of Electrical and Electronics Engineers* |
| **DSI** | *Dynamic Solution Innovators* |

# Chapter 1

# Introduction

In software development, it is a common practice for software developers to use the existing Application Programming Interface (API) libraries and frameworks. While coding for an API library, developers often get stuck which object needs to be instantiated and which sequence of methods need to be called for a particular task? To get the solution, developers frequently look for the examples and documentation provided by the vendor of API libraries, Language Forums, textbooks and unofficial websites like blogs. However, a vast number of example API usage scenarios, that are embedded in the billions of lines of already developed code are largely unexploited. That could be a very good source of repository.

Software Development has been getting challenging day by day due to rapid changeability of the type and pattern of Application Programming Interfaces (APIs). To accelerate the development process and ensure rapid productivity, software developers often reuse the code libraries and frameworks through the particular programming interfaces. All we see that, the usage information and documentation about the APIs are often incomplete and out of date. As a result, both the novice and expert developers have to face several challenges to learn new APIs. Recently, a field study of API learning obstacles was done over 440 professional developers of Microsoft Incorporation. According to the opinions and experiences of those developers, five important factors are to be considered. These are respectively designing API documentation: documentation of intent; code examples; matching of APIs with scenarios; penetrability of API; and format and presentation. Among those factors matching APIs with scenarios is the most desired one to developer's community [10].

The application of data mining technique has great advantages and potentials in developing software. Software developers often need searching existing project repositories. Code searching tool is one of the existing approaches that can guide developers by providing related code snippets and patterns. There are a number of efforts found that influences the software development. There are some existing effort like PR Miner [1], Mining Repositories[2], Perracotta [3], MAPO [4][5], XSnippet [6], Mining API Patterns [7], PARSEWeb [8], MAC[9], Scenario Based API Recommendation System[10] and etc to accelerate software development process. As we have known that Mining API usages from Open Source Repositories (MAPO) [4] [5] was one of the first and MAC[9] is one of the updated triumph to mine API usage patterns.

SpotWeb was one of the earliest attempts to help developer. It is a code search based tool. It is able to mine code examples. Basically, SpotWeb proposes a method called ColdSpots that detects the rarely used API from a given project.

XSnippet is an approach based on context-sensitive code assistant tool. It finds the relevant code fragments according to the given query. It has graph-based code mining algorithm to support the range of queries.

MAPO which stands for Mining API Usages from Open Source Repositories is one of the popular approaches in this field. It is able to identify call patterns from API usages of an existing project. It works on a query that describes a method, class, package for a particular application programming interface.

MACs mines API code snippets for code reuse. It forms a transaction database. Then a pattern database is formed from the transaction database. After an initial program statement is given, MAC is able to predict useful related API code snippets according to the initial statements. Thus it guides developers through related API usage patterns.

The existing attempts in this area like MAPO and MAC can help developers by providing call patterns and related code snippets. But to ease the development we have developed another tool which we call ESDP tool. It is able to guide the frequent API that is used in the project. We will suggest a list of APIs according to the query that is similar to the relevant API list in the databases. For example if a developer types *com.java.connection,* then ESDP tool will suggest frequents API sequences started with *com.java.connection*. So, according to the search the result of the suggestionscan be like*com.java.connection.mysql*, *com.java.connection.session*,

*com.java.connection.httpRequest* ,com.java.connection.httpResponse. And the given API suggestion will maintain the sequence. If API item I3 is used after I1 more frequently then the suggestion will be *<I1, I3>* instead of *<I1, I2>*.

When MAC predicts related API code snippets and MAPO identifies call patterns then ESDP suggests relevant and frequent API sequences. So the basic differences that we have found between existing approaches and ESDP are analyzing API similarity. MAPO and MAC compare call pattern and related code snippets according to the similarity inside the APIs but ESDP guides the developers where we match the API's similarity before entering into the APIs. In case of previous approach there is a probability that the developers are guided by showing call patterns and code snippets that is not needed to the developers.

So far, first of all we should know about software repository. Anything that leaves a trial about software development, maintenance activities, software artifact, version control system (CVS) and Bug Tracking System (BTS) or public communication tools is called Software Repository (SR). Then we also should know about mining software repository (MSR). There are a number of definitions available to define MSR. But one of the common definitions is mining software repository analyzes the rich data available in the software repositories to uncover interesting actionable information about software systems and projects. Fig. 1.1 shows a sample transformation from software source code to software repository file.

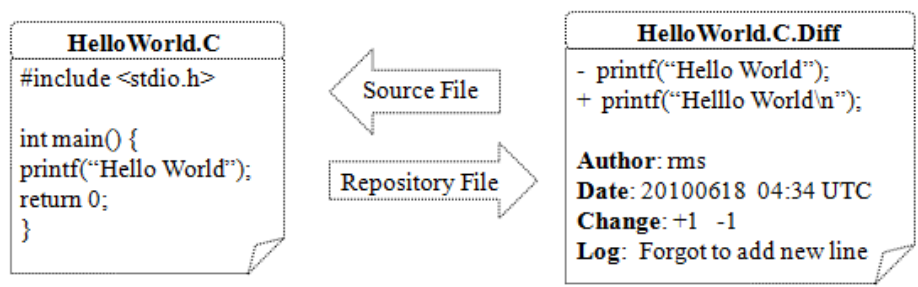

Fig. 1.1: Software source code to source repository

In our thesis we have proposed a way of providing the developers' need while they develop software. The idea we call Enhancing Software Development Process (ESDP). In general a system that speed up the development process by recommending relevant thing like API item or sequence, usage pattern, objection usage pattern etc is called ESDP as per our consideration. That means we want to guide the developers to accelerate the development process.

As we see that Developers often face difficulties while using conventional code search engines and Existing Tools like Google Code Search[16], Koders.com[26] etc. There are a number of difficulties we have pointed out that the developers face. Some of them are

i. Request and Response Latency Problems
ii. Search result is often large and it hs to deal with extra-large files
iii. Mining strategy is not that good
iv. Finding the exact match is difficult from the recommendations.

For example let's consider a scenario where a developer is writing some codes. He is writing a class called *SearchTest* as shown in the following figure 1.2.

```
1  public class SearchTest
2  {
3      private ASTParser parser;
4      private compilationUnit cu;
5  
6      protected CompilationUnit parse (ICokpilationUnit lwUnit)
7      {
8          Parse = ASTParser.newParser(AST.JLS3); //User Query
```

Fig 1.2 Developer needs Recommendations while he is coding

Now if the developers search on Google Code Search or Koders.com then he will be given a chunk of recommendations as shown in the figure 1.3 and 1.4 that makes him confused which one he might be choosing. Another thing he must be online to do so. In that case he has to face request and response latency that might consume some times.

Instead of searching on code search engines if he is more likely to search using existing code assistance tools like MAC then line he has typed will be taken as user query. The queries will be sent to Koders.com as the authors of the MAC [9] have proposed. In accordance with the relevance of the user's search query association or sequential mining will be happened. In mining two things must be held. Firstly, forming transaction database (TD) from the search results and according to the transaction database a Pattern Database (PD) will be formed by using data mining algorithm either association rule or sequential mining as MAC [9] says.

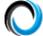

Fig. 1.3: A Search sample KODERS

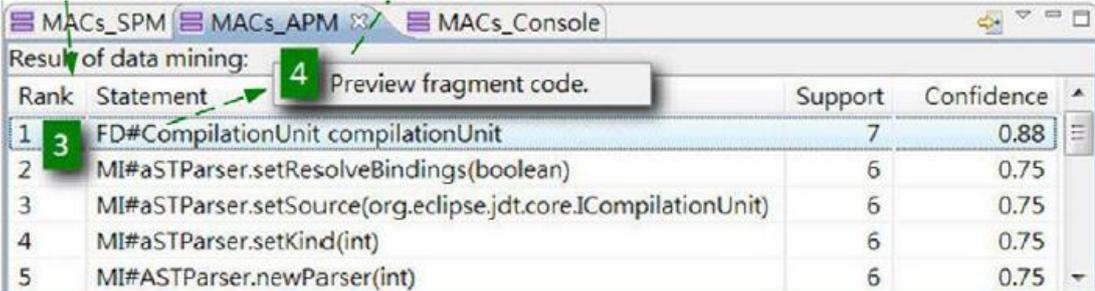

Fig. 1.4 Search Example given by Koders.com using Open Hub Black Dock

Fig. 1.5 Example of the Recommendations from the MAC

Then the developer will be given the recommendations as per that user query from the instantly mined repository. Here, the sample MAC Recommendation is shown in the following figure 1.5.

But instantly mining takes times. Developer should not be waiting for that. And the searching on the Koders.com has to resolve request and response latency. The most important thing is that they are extracting source code to an abstract form before mining. If each of the source abstraction depends on class base mining then the abstracted source item will be unique. And the uniqueness of the item will keep us away from getting relevant recommendations. Beside these there are some other issues like server dependency that really can create problem while the developer has been working on.

So from the investigation as we have ever made that comes to a decision is that an easy way of code abstraction through XML based repository in the client side can ease the development process. Advancement of a system that is able to provide a data mining integrated environment (DMIE) can fulfill the developer's demand.

## 1.1 PROBLEM STATEMENT

So we can say, we have worked through our research "*to design and develop a system with better mining strategy following special representation of enriched repository in the client side*".

We have applied four different approaches each of which is called heuristics for developing new project through ESDP framework. We have mined several open source projects to evaluate efficiency of three different tools. In the result section we have observed that ESDP provides comparatively better performances.

ESDP has its own framework that will work as a data mining integrated environment. Using framework like that will form reference snippets for the expected class, method and their pattern. Our ESDP system can be seen as an approach for the flawless, light weight, rapid software development. As per our motivation the proposed concept is different from most other existing concept as it is able to process more statement types, it not only allows for the search but also generate relevant code skeleton or sample snippets and it has an ability to provide useful API-sequences through sequence and graph mining for the rapid software development.

## 1.2 AIMS AND OBJECTIVES

Enhancing Software Development Process (ESDP) is precisely a goal oriented research. It has multiple purposes to be fulfilled. Objectives with specific aims are prescribed as in the following-

- Developing an easy and less complicated way of code abstraction through XML file based repository system without using database server.
- To speed up software development process by reducing amount of time in the coding stage with the help of set of mined existing projects.
- Advancing a system able to provide data mining integrated environment to the programmers.
- Analyzing the comparison of efficiency of the existing concept with the proposed concepts.
- Drawing conclusion showing future the prospects of the idea of "Enhancing Software Development Process (ESDP) using Data Mining Integrated Environment".

From the real world observation we have come to see that Dot Net Framework is documented very well unlike others. It provides typical snippets along with other documentations. But in case of many API, the provided snippets show only single use instead of multiple applications. As we know a method or class in the API might have multiple uses so the usage of these may not be exhibit relevantly to the task we need to do. That is why the programmers often see that the associated documentations provided by an API builder company are merely useful.

Code search engine like Google Code Search [11] or code snippet recommenders like Strathcona [12] can be a helping hand for the novice programmers towards finding a solution of the problem they face. As we know open source projects have different usage patterns, so applying search engine or snippet recommenders can give useful results after searching through the files. If the open source API libraries are huge then the search results found are also huge. Ranking or indexing the search result is another challenging task for the programmers. Existing tools like Google Code Search, Strathcona, MAPO [4] [5][13] have already proved their weakness in the code searching and browsing issues.

So from the finding we have ever investigated to find the answer of different unanswered questions.

Firstly, what are the issues should be considered before mining the repository. How an enriched repsitory influences the search matching?

Then how we can represent the code abstraction using *XML file based repository* system instead of server based data manipulation system like database system?

Thirdly, how we can get facilities from the client based repository with mining instead of web based mined repository. How server dependency influences the recommendations of the software repository mining process.

The last research question we have tried to find the answer is how to increase the efficiency of the recommendations through graph mining algorithm that is applied on the source skeleton.

## 1.3 COMPARATIVE CONTRIBUTION

Data mining [14] gives different useful techniques to mine huge scale of data. The techniques in some cases have some drawbacks but in our work titled "Enhancing Software Development Process (ESDP) using Data Mining Integrated Environment" is able to bring a better outcome to help the programmers. ESDP-tool is able to provide necessary usage pattern after traversing the existing library files mined to it. With the help of that useful usage patterns ESDP-tool further help the programmers. ***Weclaim that our work has made the following contributions-***

- **Enriched and Updated Repository Mining Strategy:** Repository building is an important issue in Mining Software Repository (MSR). If the source repository is more updated and enriched then the mined system will be easily capable of to recommend relevant suggestions and patterns. There are a number of issues should be considered before building a repository. We have considered multiple important sources of projects to build our central repository. Besides enlarging the sources of codes we also have considered that how often the repository is updated to ensure the contemporary need of the developers.

- **Source Abstraction Technique using XML Based System:** A special type of techniques we have applied to extract the API usage pattern information and documentations from the source code files. In different existing approaches a distinguish code analyzer is able to analysis the code snippets including API classes and methods. But in case of our ESDP tool we have applied kind of different technique to extract the raw source code to an intermediate form of source abstraction. First we extract these to code readable files. Then we translate the codes to an abstract form. Detailed source code is the appropriate way to represent the behavior of a system but the detailed format itself has some limited applicability for analysis and mining purposes.

- **Server Independent Searching and Recommendation Procedures:** ESDP tool provides a particular user interface to suggest usage patterns and their related snippets to search the necessary snippets inside the mined repository. Unlike other approaches we have stored our mined API into XML based repository in the client side. So the user can easily search and get recommendation according to their user query they type. To avoid server concurrency and latency problem we have considered client based repository instead of online based database system.

- **Usage of Graph Mining Algorithm:** We have applied graph mining algorithm called *pattern explorer* to enrich the recommendation. In our approach the usage of a set of objects in a scenario is transformed code skeleton to a Directed Acyclic Graph (DAG) of which nodes represent constructor calls, method calls, field access and branching point of control structure and edges represent temporal usage order and data dependencies among them. An individual usage pattern is a sub graph that frequently appears in mining object usage graphs. Thus we detect the usage patterns related to the code snippets and recommend to the developers.

# Chapter 2

# Background and Related Work

In software repository mining there are a number of existing approaches. Some approaches have better recommendation system and some approaches have enriched repository. If the repository is enriched enough then we will get better recommendation. There are some approaches that work in association with online based code search engine like Koders and Google Code Search. And the search results found in response to a particular query is then mined before the final recommendation is given to the developers. Here we have explained different existing approaches that are widely used inMining Software Repository (MSR).

## 2.1 APPROACH OF MINING TO CODE REUSE

MAC [9] is an approach to mine API code snippets for code reuse. They formed a transaction database. According to that transaction database they developed pattern database. And thus MAC suggests by evaluating the support, confidence and rank list of the frequent item.

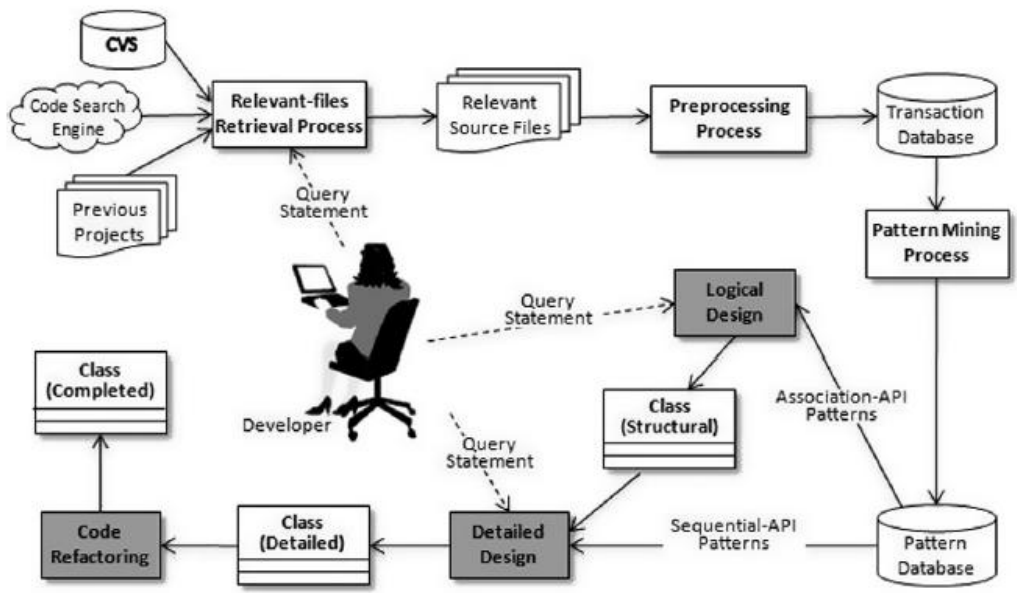

Fig. 2.2: Mining API Code snippets for code reuse

We also have investigated the MAC [9] framework as the following figure 2.1shows. MAC is of the updated approach that is able guide to developers through association and sequence mining algorithms.

Mining API Usages from Open Source Repositories (MAPO) [4][5] is one of the popular approaches in this area. It identifies the call patterns from API usages. Ausagequery is that describes a method, class, package for an API. They use koders.com as their search repository. Fig. 1.2 shows the MAPO framework.

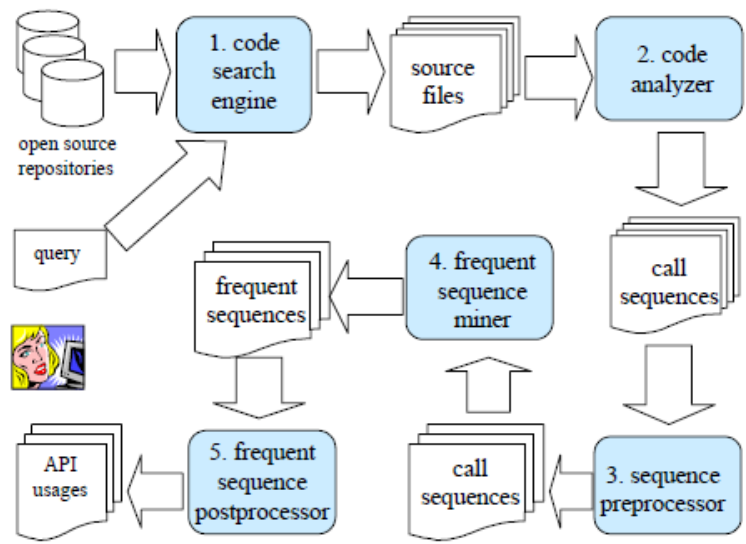

Fig. 3.2Mining API Usages from Open Source Repositories (MAPO) Approach

Another approach Strathcona [11] gives a number of relevant snippets by matching the structure of the code under development with the snippets belongs to the repository.

CodeBroker [12] is mostly similar tool to Strathcona. It automatically searches the repository by using comments provided by the developer. CodeFinder [13] uses a query browser to help the developer construct queries that can be sent to the repository.

XSnippet [6] was developed by Tansalarak and Claypool. They extend Prospector and add additional queries, ranking heuristics, and mining algorithms to query a code snippet repository for the relevant snippet at hand.PARSEWeb [8] developed by Thummalapenta and Xie used Google code search for collecting relevant code snippets and mines the returned code snippets to find the solution. Saul proposed an approach [17] to find API methods that are closely related to a query API method of interest, by discovering API methods that share a caller or a callee with the query API method.

Another attempt is GrouMiner [18] [19] a novel graph-based approach for mining the usage patterns of one or multiple objects. GrouMiner approach includes a graph-based representation for the multiple object usages, a pattern mining algorithm and an anomaly detection technique that are efficient, accurate and resilient to the software changes.

## 2.2 APPROACH OF MINING TO ANALYZE THE SYSTEM

Another tool that automatically builds queries to send to the repository is the Hipikat tool [14]. Hipikat creates links between different sources of information in a project, including source files, cvs commits, bug reports, newsgroup postings, and web articles.

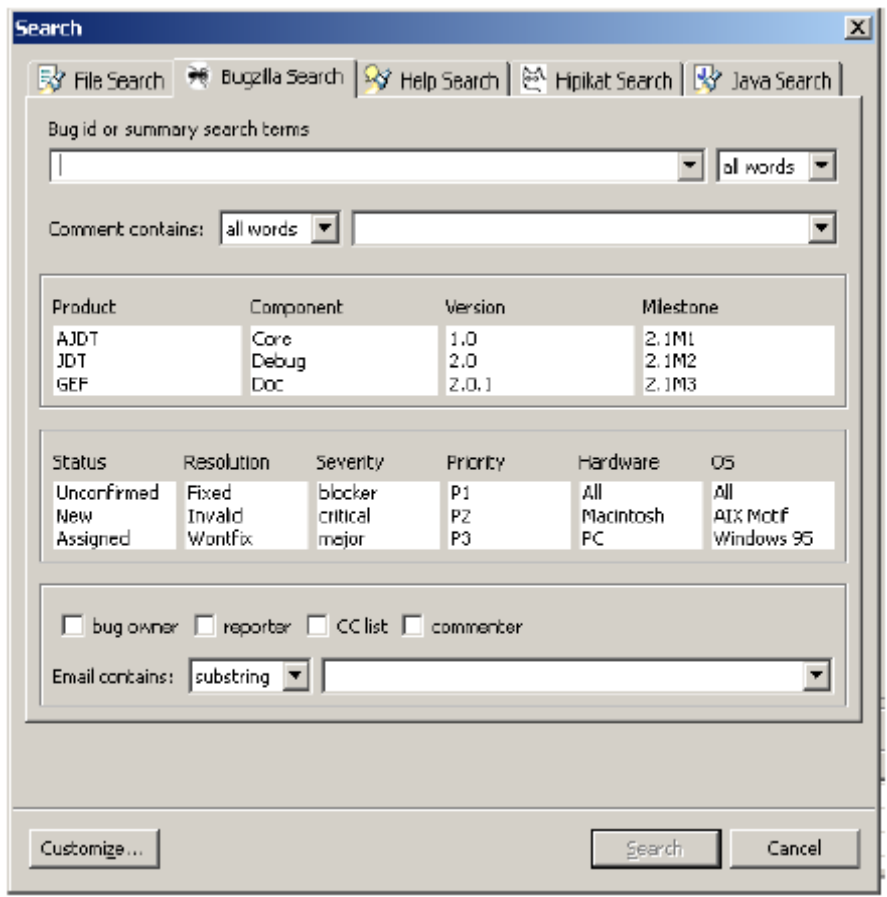

Fig. 2.4: Hipikat Tool

Hipikat is a tool intended to solve this problem. Hipikat recommends relevant software development artifacts based on the context in which a developer requests help from Hipikat.Hipikat is an ongoing research project. To investigate our ideas, we are developing Hipikat as an Eclipse plug-in to support development of the Eclipse integrated development environment platform. Please be aware that this working prototype recommends only in the context of Eclipse development (that is, it knows only about artifacts in *dev.eclipse.org:/home/eclipse* CVS repository, *eclipse.org* Web site, *bugs.eclipse.orgBugzilla* and *eclipse.tools* newsgroup). In particular, Hipikat does not know about any other projects you might have in your workspace. One of the working screen shots is shown in the fig. 2.3.

The Reuse View Matcher [15] uses a repository of constructed examples to demonstrate how individual classes in the framework can be used. As these examples are constructed by the framework-authors, their correctness and applicability can be more assured. However, this approach suffers from the fact that creating the examples is time consuming and their coverage cannot be completed.

Mandelin initiated an important effort in this arena called Prospector [16]. It tried to synthesize solution jungloids from jungloid query.

## 2.3 APPROACH OF MINING TO GUIDE SYSTEM CHANGES

Rose developed by Zimmermann can guide programmers to locate possible changes. Historical information, such as source code and related files, is used to suggest likely changes. This information prevents errors due to incomplete changes and finds couplings undetectable by program analysis. Moreover, Ying, Murphy, Ng, and Chu-Carroll and Hassan and Holt also show that suggestions based on historical co-changes that are useful to correctly propose the entities which must co-change.

## 2.4 APPROACH OF MINING TO PREDICT AND DETECT BUGS

An approach initiated by Williams and Hollingsworth used the source code change history of a software project to search for bugs. Their study shows that their bug-finding technique is more effective than the same static analysis that does not use historical data from the source code repository. Another contribution in this arena was done by Li, Lu, Myagmar, and Zhou by proposing a tool called CP-Miner, that uses data mining techniques to efficiently identify copy-pasted code in large software suites and to detect copy-pasted bugs. Several additional studies provide extensive discussions of the applications of data mining to the field of detecting and predicting software bugs.

## 2.5 APPROACH OF IMPROVING USER EXPERIENCE

An approach to improve to user experience is Stabilizer by Michail and Xie in 2005. It allows users to collectively help each other to avoid bugs especially in GUI (Graphical User Interface) applications. The tool can mine reported bugs and execution logs to prevent an application from crashing. When users attempt actions that have led to problems in the past, they will receive a warning and be given the opportunity to abort the action. Another unified framework given by Mockus, Ping, and Li was developed to investigate and predict the effort, schedule, and defects of a software project. Instead of mining the source code data, it mines data captured by project monitoring and tracking infrastructures, as well as customer support records, to determine the expected quality of a software application. They also created tools to retrieve, process, and model such data to understand the relationships among process and product factors and key outcomes, such as quality, effort, and interval. They found that the deployment schedule, hardware configurations, and software platform have a significant effect on the perceived quality of an application, increasing the probability of a software failure by more than 20 times [9].

The previous approaches mine properties without temporal information, whereas in our ESDP tool we will mine more complex patterns involving multiple methods and temporal information.

## 2. 6 PROBLEM FINDINGS IN THE EXISTING RELATEDAPPROACHES

In the previous approaches there are some flaws as we have investigated. Some approaches build their repository without regular update concepts, some approaches search query from the server based repository like database or web based repository [4] [5] [9]. There are some approaches where mining is held during the instant query execution [9]. There are also some approaches found that works only graph mining strategy without integrating the general mining approach to the graph mining algorithm [18] [19]. These things work but in some cases it can bring hassles to the developers. Representation and abstraction of the source code is a vital thing in source repository mining. If the representation is not that readable to the developers then the recommendations does not make any important sense. Server dependency of different software repository mining tools is another issue. If everything belongs to the remote server then request and response might consume valuable times of the developers. So it will bring a better outcome to the developersif the framework of the repository tool has some distinguished characteristics as mentioned below.

i. Enriched and Updated Repository
ii. Better Source Abstraction
iii. Convenient Source Representation
iv. Easy and Smooth Mining Strategy
v. Server Independency
vi. Integration of Graph mining algorithm.

If an approach can integrate these issues in the single framework of the repository mining tool then the developer will be highly benefited. In our ESDP system we have tried our best to do so.The Mining Approach of our ESDP differs from the Existing Approaches. In the Existing Approach[9] they mine the API item where each item is unique as they have taken the location of the item like line number. So it is difficult to find the resembling item While mining because each of the item becomes unique if that type of class based mining is used. Another thing is that as most of the existing approaches do not have their own code searching tool instead they depend on some online based code searching tool like Google Code search, KODERS and etc. In that case, request and response latency and search dimension become a deciding issue that surely influence the efficiency of getting relevant matches within expected time. In our approach we have taken the particular part of the API item instead of taking the location or line number like. Beside, we have used XML based representation in the client site to avoid server interaction instead of conventional database dependency.

# Chapter 3

# Proposed Idea and ESDP Framework

It will be a better software repository mining system if the system has some special features integrated. These special features are like better representation, server independency, enriched repository, fast and accurate recommendation etc. For example before store the source code lies in the repository needs to translate from the detail source code into an abstract form. The process needs to do so may vary approach to approach. In our system we have a different representation system. Another vital issue is server independency. In our ESDP system we have designed the system without server dependency. Even for storing the mined sources we have applied XML based system. That is why it is able to handle multiple search queries simultaneously without any concurrency and latency problem. Thus our ESDP system is kept away from the server failure problem. The ESDP framework is shown in the figure 3.1.

To extract usage pattern from the given code segments we will apply data mining technique. We will filter the source files before applying mining procedure over the set of referenced codes available. Then we will need to check different combinations among code pattern.

According to the particular combination we will decide which of the technique can be applied. If the existing API is small in size then ESDP will apply Apriori principle [20] to find the frequent item set. If the API files are comparatively large according to the given threshold file size then another algorithm FP-Growth [21] will be applied. And to find the frequent sequence pattern we will use PrefixSpan [20] algorithm. The API files will be inserted into XML repository just as usual tables. The following steps will be applied before we suggest the usage pattern and their snippet to the novice developers.

To automatically mine Classes and APIs usage pattern from the existing source repositories we will design a framework. The framework will receive query describing the name of method, class or package and will give output a set of API usages and snippet.

We have applied our own idea to provide better outcome comparing with the existing concepts. For that we have a distinguish framework as shown in the above figure. In the framework as shown in the figure 3.1, there are four different parts that work together. These are explained here below by turn.

## 3.1 REPOSITORY HEURISTICS

Repository building is an important part in software repository mining. If the source repository is more updated and enriched then the mined system will be easily capable to recommend relevant suggestions and patterns. There are a number of issues should be considered before building a repository. Two of these are very fundamental that are listed below.

i. What are the sources of codes that are used to build the repository?
ii. How often the repository is updated?

First of all we have thought about the fundamental sources of the repository. If the sources are limited then the mined repository will be comparatively light weighted. We have investigated on different data mining based on API recommendation system like MAPO [4] [5], GROUMINER [22] [23] and MAC s[9].

In MAPO [4] [5], the sources are taken from open source repositories. They have not used any other sources like regular update, external APIs, standard libraries as their sources of repository.

GROUMINER [22] [23] is one of the popular research works in APIs recommendation system from mind repository. They also have built repository from open source projects available online.

MACs is another one of the contemporarily published research work. They have mined sources according to the user query instantly given by the user. Before mining they have dynamically built the repository using Koders.com.

In the above mentioned works they have not considered standard library APIs of a particular platform as well as APIs found from the regular searches or from external APIs.

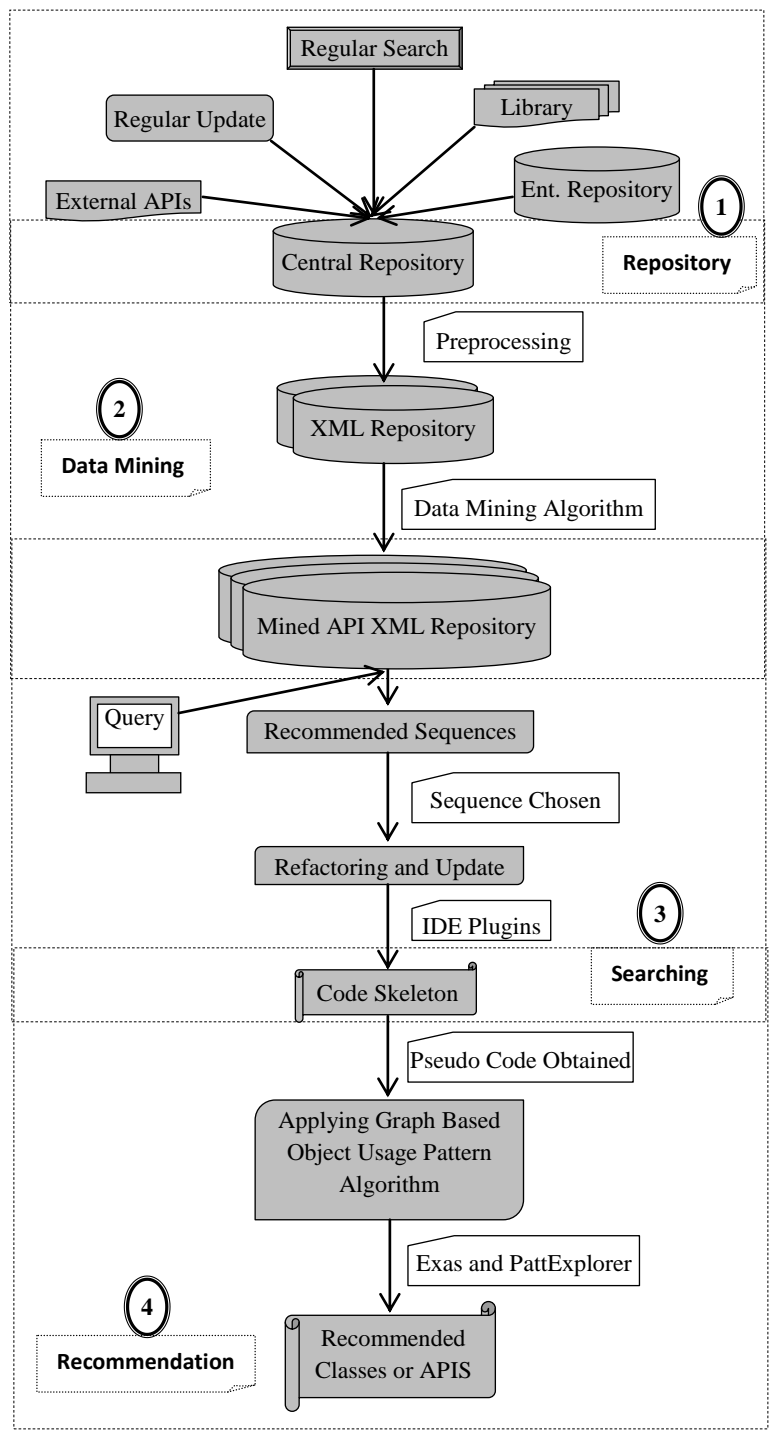

Fig.3.1 Proposed ESDP Framework

If the open source projects and the code search engines like Koders.com or Google code search engine become the only source of building repository then the mined repository will be obsolete and expired after certain duration of time.

Another thing is that real time software development is getting changed so rapidly. The developers today have been using a set of APIs or framework that might be old tomorrow. So prediction is merely applicable on what will come tomorrow with the development change and evolution. We should open up the door of regular update to keep ourselves sustainable and cope with the modern challenges in rapidly changing software industry.

Before we work on our ESDP system we have considered these challenges. In ESDP the central repository is formed from different types of dynamic sources. We have done it to keep our repository up to date.

As we will see that, we have five different sources of repository that provides APIs and class sources to form central repository of the repository heuristics of ESDP tool as shown in figure 3.2. The sources are-

    i. Enterprise Repository
    ii. Standard Library
    iii. Regular Searches
    iv. Regular Update
    v. External APIs

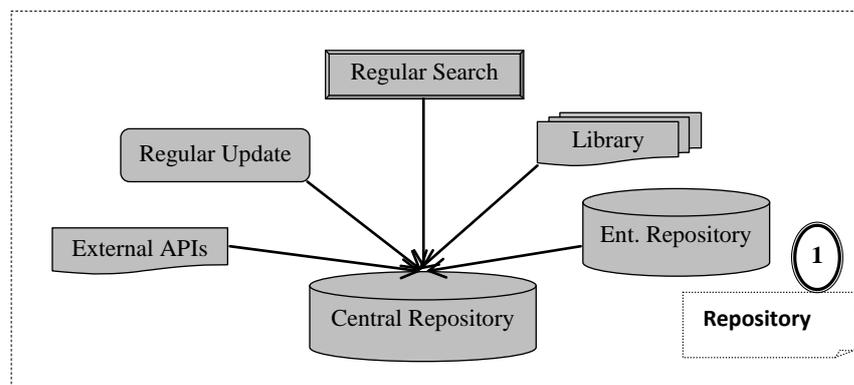

Fig.3.2 Repository Heuristics of ESDP System

### 3.1.1 ENTERPRISE REPOSITORY

Software is developed by a particular company using a platform. After certain duration the company might have a number of projects that follows the same platform and framework. Let's consider a company that has developed a project which contains similar type of APIs or code pattern. The development of the proposed software might ease the process and save time if the company gets suggestion and recommendations from the previous completed project. Considering this phenomenon we have contacted with a leading software company [24] in Bangladesh about taking their sources repository. They have conditionally agreed to provide some projects sources only purposing research. Thus enterprise repository is one of our sources of building central repository.

### 3.1.2 STANDARD LIBRARY

Suppose a novice developer is going to develop project using a particular platform java. So java standard library and documentation could be an important source of building central repository. For example if this developer needs collection APIs then he or she need not to seek for somewhere else away from the java standard collection APIs. If he execute a search query alongside standard APIs they he will be given number of suggestions and recommendations in response to that search query. Considering the easiness and well decorated features of the standard library of different platforms we have used standard APIs as another source of building central repository.

### 3.1.3 REGULAR SEARCH

We have seen that MAC [9] has used a code search engine to get the desired item following a search query. In case of MAC, searching happened before mining instantly. But if we bind the search dimension only inside the area of code search engine like Google Code Search Engine or Koders.com then it comparatively keeps us away from the exact match. Searching rapidly online using a user query has some drawbacks itself. For example let's consider that a developer is typing his code on an IDE editor. All on a sudden he seeks for a specific term or item and he starts searching in the internet. After he gets the desired result he continues working and after a couple of minutes he looks for another item again. Thus he does the same as he has done just before. Unfortunately if it takes longer time due to latency and concurrency issue. That means anyway it kills some valuable time of the programmer. So we see that rapidly searching on the web during development period is not that convenient because of server dependency. Server dependency has different other issues to resolve. For example security issues, connection issues etc. How much a developer is likely to get connected to respond with an unknown server while he is on the version control system that does matter? We have seen the comparison between the response time of the server and client. It is easier and faster to search from the client offline than searching from somewhere dependent on the internet.

Considering this issues and draw backs all we have done is that we have gone through and tracks the trending search terms of a code search engine and get connected into our tools to eliminate server dependency. According to the search queries we have added the prominent and frequently used APIs to our control repository. This is also mentionable that after an interval we update our central repository according to the trending search terms found right then mine the code searching protocol.

3.1.4 REGULAR UPDATE

All we see in the previous works that the repository is not updated after it is built once. Even MAC [9], MAPO [4][5], GROUMINER [22][23] has not explicitly mentioned the updating concept. If the repository is not updated then the mined repository eventually fails to fulfill the user required matching. Considering the updating issues of the repository we keep our central repository updated after three months of interval.

3.1.5 EXTERNAL APIS

Besides enriching our central repository from multiple sources we have open up an opportunity to add API files from the external sources. As we see that the new APIs comes with times. We have some APIs writer that we call ESDP API

Developer. ESDP API developer will write newer APIs to survive and sustain with the development pattern change.

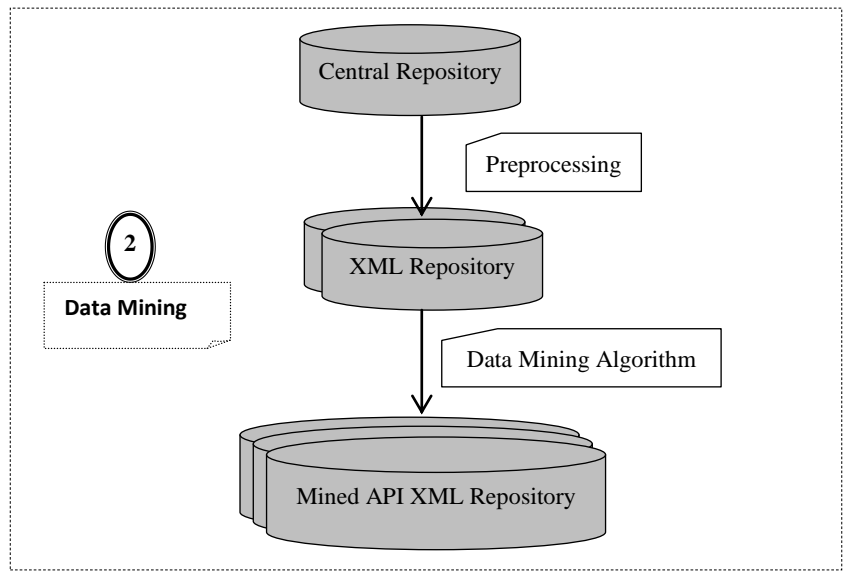

Fig. 3.3 Data Mining Heuristics of ESDP Framework

## 3.2 DATA MINING HEURISTICS

The second heuristics of ESDP Framework is captioned as Data Mining Heuristics. Here, basically two things happen. In the first step the central repository is preprocessed to an XML Repository following a special type of XML conversion strategy. Then in the last step a data mining is applied on the XML repository to build mined API XML repository as shown in the figure 3.3.

### 3.2.1 DATA PREPROCESSING

In the central repository the API and class files are stored as .java or .jar files. First we extract these to code readable files. Then we translate the codes to an abstract form. Question may arise that why we don't use detailed source code. Yes, detailed source code is the appropriate way to represent the behavior of a system but the detailed format itself has some limited applicability for analysis and mining purposes [9]. Let's consider the two field declaration statements in java.

*Connection connection;*
*Connection conn;*

Here, the object declaration is made with just different names. The two statements have the same meaning. To overcome this problem, we translate the raw source code into a higher level of abstraction form. The common item form was expressed as shown in the following table 3.1 and 3.2.

Table 3.1 Source Abstraction

| Serial | Generic Example | Real Example |
|---|---|---|
| 01 | type, name, entity, location | FD, dom.ASTParser, com.Test:05 |
| 02 | type, name, entity, location | MI, method_A(java.lang.String): void, com.class_B.method_C():130 |

Table 3.2 Details of Source Abstraction

| Key Term | Elaboration |
|---|---|
| FD | Implies for the Field Declaration. FD is an item. |
| dom.ASTParser | Field Declaration (FD) statement. dom.ASTParser is an item name. |
| Com.Test:05 | appears at the line 5 within class Test, of package com. |

In the second real example [Table 3.1 and 3.2] a method invocation *method_A()* with a parameter in type of *java.lang.String* and return void type appearing in the method *method_C()* that is inside the class *Class_B()* which is under the package *com*.

In our ESDP tool, we have considered 17 types of items purposing our ESDP research as shown in the Table 3.4. Our approach can be applied for any object oriented programming language, but for the ease and availability of evaluation and its correctness the considered platform here is Java 2.0.

3.2.2 SOURCE CODE TO XML TRANSLATION

In this step we cover the abstraction code with the XML meta tag. An example is given below.The above transaction represents the field declaration *javax.swing. JButton* of the class *classB* in the package *pkga* and its position (line number) inside the file is 87 as shown in figure of temp1 in the bold tags. A transaction is the set of entities simultaneously used in a block such as class block or method block. A sample set of entities is given in the following equation 3.1. The element of the XML transaction is called item. Each item represents a way to use one or more specific API and these are the basis for later mining process.

Table 3.1 Considered 17 types of items purposing our ESDP research

| Serial | Code Item | Description | Example |
|---|---|---|---|
| 1 | PD | Package Declaration | package foo.biz; |
| 2 | ID | Import Declaration | import javaioFile; |
| 3 | TD | Type Declaration | public class Example_Class {}' |
| 4 | FD | Field Declaration | private Connection conn; |
| 5 | CI | Class Instance Creation | newFile (); |
| 6 | MD | Method Declaration | public File method (String s) { } |
| 7 | MI | Method Invocation | method.open (filename); |
| 8 | VD | Interface Implementation | implements Runnable { } |
| 9 | IF | Local Variable Declaration | String s |
| 10 | ACD | Anonymous Class Declaration | new Enumeration () { }; |
| 11 | AA | Array Access | file array[i]++; |
| 12 | AC | Array Creation | newFile[n]; |
| 13 | CTI | Constructor Invocation | this (parameter); |
| 14 | FA | Field Access | object.field_A = 2; |
| 15 | SCI | Super Constructor Invocation | super (parameter); |
| 16 | RT | Return Statement | return a_file object; |
| 17 | SC | Super Class Inheritance | extends SuperClass { } |

$$T = \begin{cases} (VD, \ldots\ldots\ldots\ldots\ldots..Core.ICompilaionUnit, p.ClassA.m(\ ):40), \\ (VD, \ldots\ldots\ldots\ldots\ldots\ldots core:dom:ASTParser; p:classA:m():97) \\ (MI, \ldots.dom:ASTParser:setKind(int); p:classA:m():155), \\ (MI, \ldots..dom:ASTParser:setSource(); p:classA:m():210), \\ \qquad\qquad\qquad\qquad . \\ \qquad\qquad\qquad\qquad . \\ \qquad\qquad\qquad\qquad . \\ \qquad\qquad\qquad\qquad . \\ (MI, \ldots dom:ASTParser:setResolve\ldots(); p:classA:m():260) \end{cases} \text{eq}^n \text{ (3.1)}$$

3.2.3 APPLYING DATA MINING ALGORITHM

The ultimate goal of API code snippet mining is to mine the sequential usage pattern (rules) from a given repository and that are relevant to the task at hand. In our thesis we have proposed and used *PrefixSpan* as sequential pattern mining algorithm.

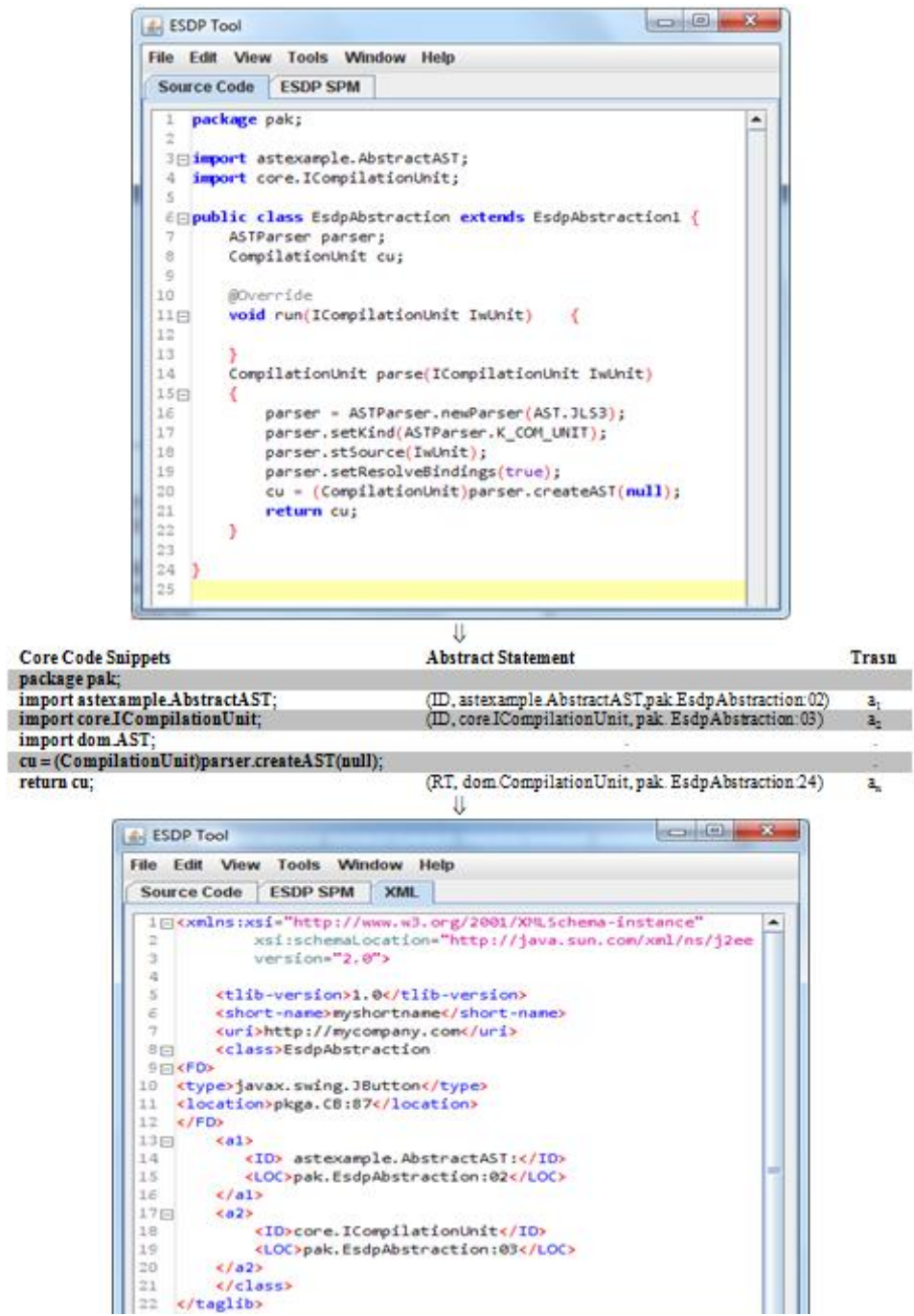

Fig.3.4 Source Abstraction from Raw Source to XML

3.2.3.1 SEQUENTIAL PATTERN MINING

The API code snippet pattern mined by association rule pattern mining can be used to create the structure of a class including class attributes and method definitions. However the association rule mining does not consider the order of the transaction. In a method block ordering of invocation is important. For example in an API calling sequence developers call the methods of an object after they declare the object. At these situations association rule mining is not appropriate, Sequential pattern mining is needed. Given the XML transaction as shown in the following figure 3.4, translated from the statements in methods in the previous sections, we can create a sequence XML files, S. The sequence XML files consist of a set of transaction along with <SID, S>, where SID is a sequence ID and S is a sequence. In ESDP the SID is an XML tag that is the location of the sequence. A sequence <SID, S> is said to contain a sequence α if α is a subsequence of S. The support of a sequence α in a sequence XML files is the number of sequence in the files containing α. It can be denoted as $support_s (α)$ if the sequence data sets are clear from the content. Given a positive integer min_support as the minimum support threshold a sequence α is frequent in the XML sequence files S if and only if $support_s (α) \geq$ min_support. That is, for sequence α to be frequent it must occur at least min_support times in S. A frequent sequence is called a sequential pattern. A sequential pattern with length K is called a k-pattern sequence. Here is the example,

$Support_S (α) = |\{< SID, S > | < SID, S > \in (S) \land (α \subseteq S )\}|$  eq$^n$ (3.2)

$A = <S_1, S_2, S_3, S_5>$  eq$^n$ (3.3)

Where, $S_1$ = MI, dom.ASTParser.newParser(int):dom.ASTParser

$S_2$, $S_3$, $S_5$ refer to the values in the figure of abstraction API forms respectively. Here, there are four instances of items in the sequence α, therefore, it has a length of four and is called a 4-sequence. We have used a recently proposed sequential pattern mining algorithm, called prefixSpan (Prefix-Projected Sequential Pattern Mining). This is a sequence mining algorithm which is a pattern growth method that does not require candidate generation. The algorithm mines the complete set of patterns, but greatly reduces the effort of candidate generation. It can reduce the projected repository size and lead to the efficient processing.

### 3.2.4 BUILDING MINED API XML REPOSITORY

We apply sequential pattern mining algorithm (PrefixSpan) to the preprocessed data results to find the API sequence pattern from the XML repository. Each XML repository consists of API-usage patterns and each pattern consists of a set of API code snippets that have been used together frequently in the past. To find the frequent API pattern we apply different queries using our referral programming language feature like String Manipulation and XML Parsing. The Querying and Ranking phase using the strategy of string manipulation and XML parsing APIs take a set of candidate recommendation, order and filter it from the set to generate a refined recommendation list. The list are ranked and covered by XML tags as shown in the figure 3.5. ESDP recommends the sequential API code snippets and each recommendation includes several statements. The amount of statements in a sequence is called k-sequence. We took the product of k and the support value of the sequence to rank the API pattern in the mined XML repository.

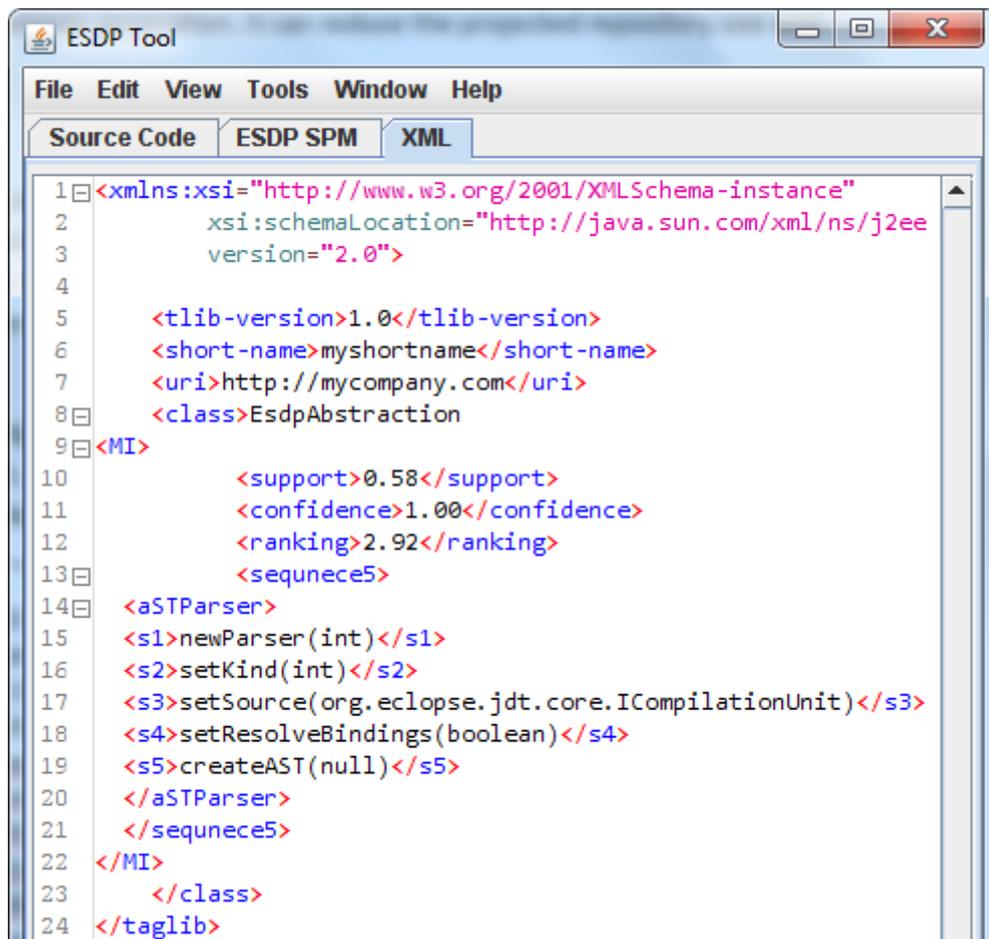

Fig.3.5 Mined XML Repository

## 3.3 SEARCHING HEURISTICS

In searching heuristics there are several steps need to be traversed to get the code skeleton. Here a sequence recommendation is given according to the user query searched by the developers. If the recommendation is chosen by the developers then it can be refactoring and updating if the developers want to do so. After doing that a code skeleton is found. This thing is done by searching heuristics. And the Searching Heuristics is shown in the figure 3.6.

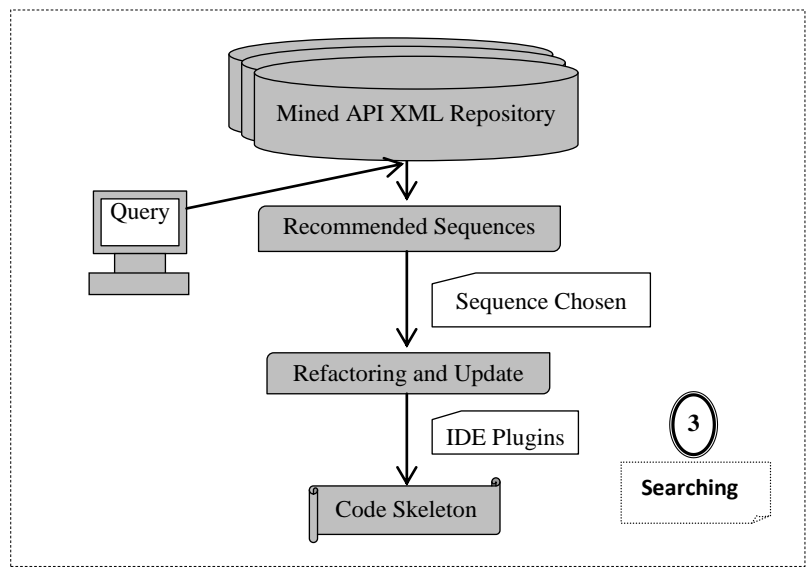

Fig.3.6 Searching Heuristics of ESDP

### 3.3.1 RECOMMENDING SEQUENCES

After the mined XML repository is built, searching is kind of easy with a particular user query statement. Here we get a set of fragment code into a method block after getting searched. As we know that the code snippets are stored in the Mined API XML repository according to their frequency, ranking, support, confidence and with necessary methods and fields' snippets. So when the developers look for a suggestion the typed statements is taken as user query and searched to find the required matching in the Mined API XML Query.

For example, let's consider that the user is writing a class called *SearchTest*. He has written some statement like below. While he is typing the line as shown as user query in the figure below he wants suggestions from the ESDP tool.

Fig.3.7 A user query is marked when the developer is writing codes

Fig.3.8 Match is stored in Mined API XML Repository

The certain statement as shown in the figure 3.7 he marks as user query will be sent to the Mined API XML Repository to find the match. As we already have mentioned that the statements and the associated methods and attributes are stored in the Mined API XML Repository with the support, confidence, rank and sequence number. The sample XML representation is shown in the figure 3.7.We get by querying the sequential pattern rules with the statement as shown in the figure 3.8.We use it as the input to query the relevant statement sequences. The example shows that we found several statements sequences ranked with their ranking scores (fig3.9). We choose the first recommendation in the form of code. Then we can update and refactor the recommended statements in the preview window (figure 3.10) of ESDP plug-in inside the IDE.Finally we get code skeleton as shown in the fig. 3.11.

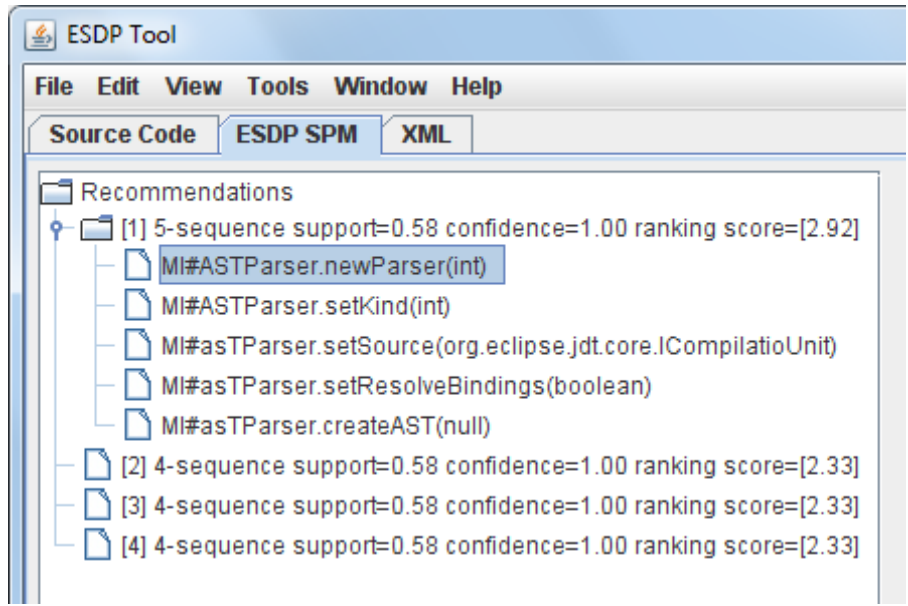

Fig.3.9 Suggestion after Sequential Pattern Mining in response to the user query

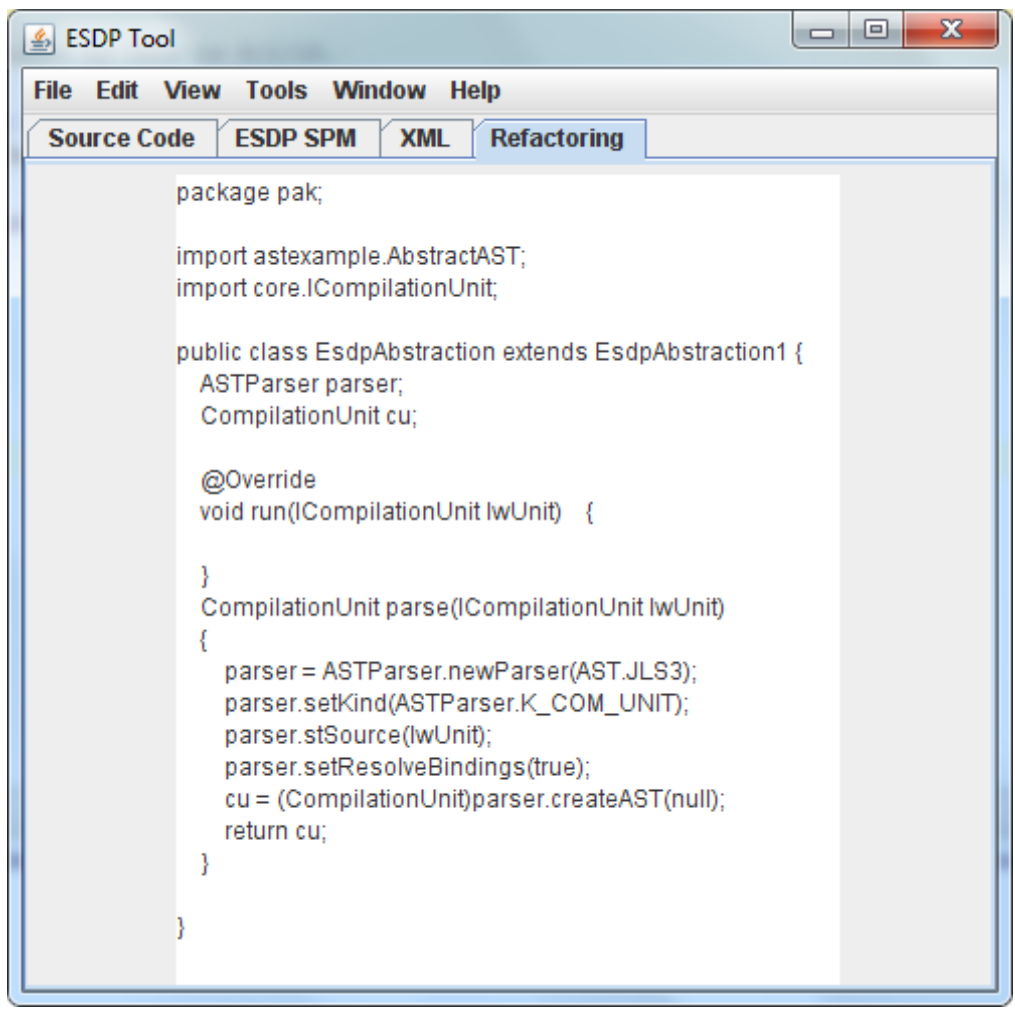

Fig. 3.10 Refactoring and Updating Window of ESDP Searching Heuristics

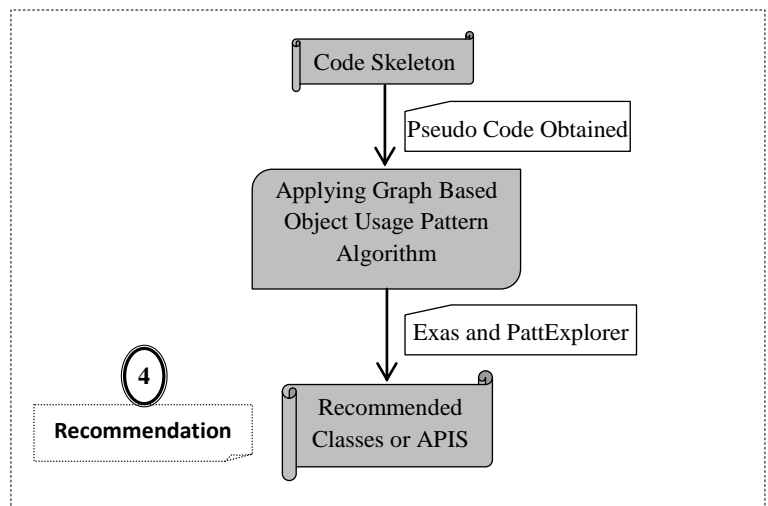

Fig.3.11 Code Skeleton

Figure 3.12: Recommendation Heuristics of ESDP

# 4. RECOMMENDATION HEURISTICS

In Recommendation Heuristics we apply another approach for mining the usage pattern of the objects and classes using graph based algorithm. In our approach the usage of a set of objects in a scenario is transformed code skeleton to a Directed Acyclic Graph (DAG) of which nodes represent constructor calls, ,method calls, field access and branching point of control structure and edges represent temporal usage order and data dependencies among them. Recommendation Heuristics is shown in the following figure 3.12.

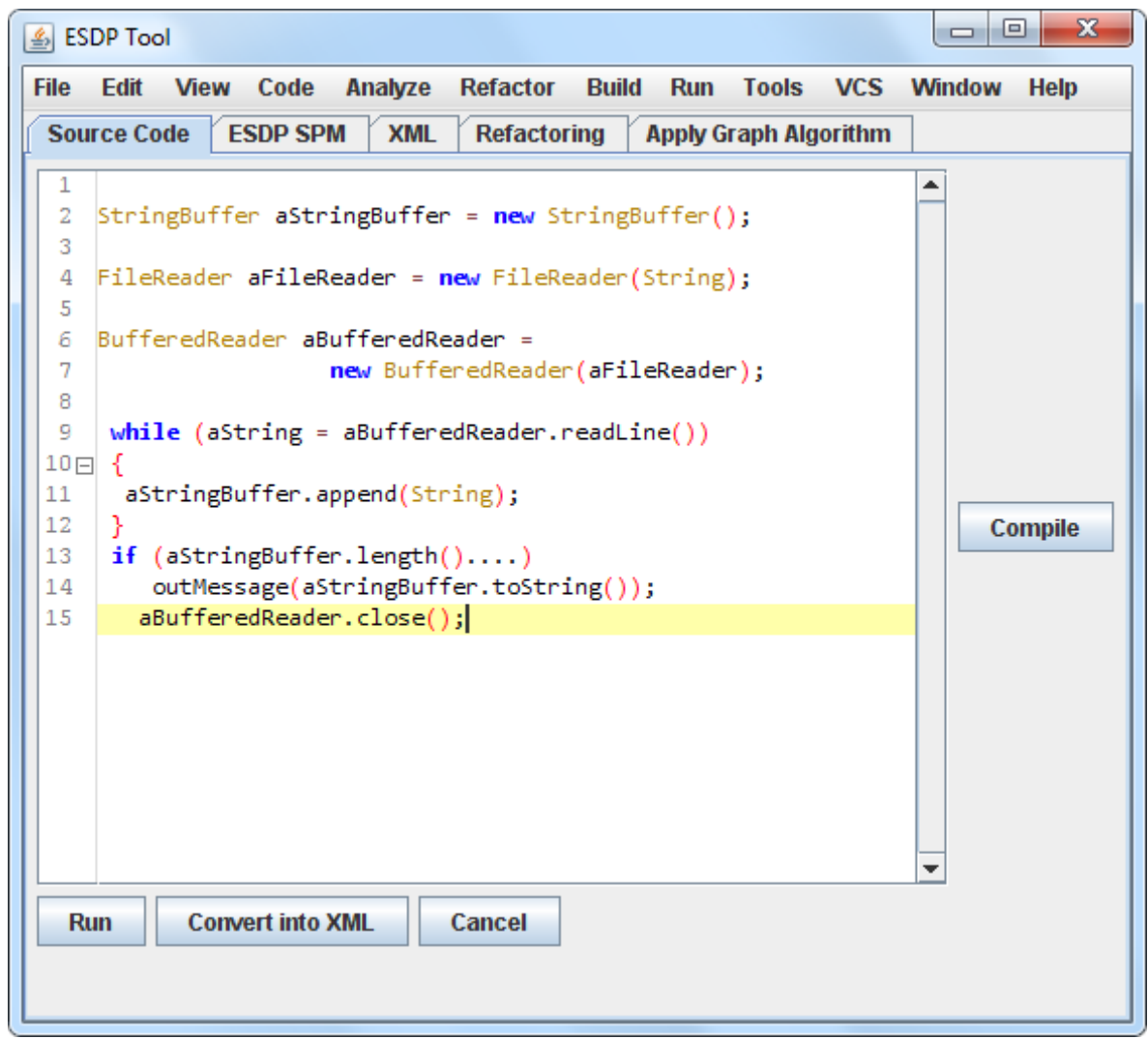

Fig.3.13: Code Skeleton

A usage pattern is considered as a sub graph that frequently appears in the object usage graphs extracted from all methods in the code base.

In Object-Oriented Paradigm, an object interacts with another object of the same or different class by invoking methods and fields. As we can see the object or method call and control flow as shown in the figure 3.13 and a 3.14 among them eventually follow specific orders or control structure constraints in considering the class or interfaces. Specific orders and/or control flows of objects' method calls cannot be checked at compile time. If development team members lack of usage documentation due to busy schedules as a consequence, errors could not be caught until testing and even go unnoticed for a long time. Inside our ESDP tool we have converted the existing source snippet to directed acyclic graph (DAG) [23] is shown in figure 3.15 where nodes represents method calls, field accesses, constructors, branching and looping of control structures and edges represent temporal usage orders and data dependencies among them.

Here, an individual usage pattern is a sub graph that frequently appears in mining object usage graphs. We detect the usage patterns using a novel graph based algorithm for mining the frequent induced sub-graph in a graph dataset. The patterns are generated increasingly by their sizes (i.e. number of nodes). Each pattern of size k+1 is discovered from pattern P of size K via extending the occurrence of P in every method's graph G in the dataset with the relevant nodes of G. The generated sub graphs are then compared to find isomorphic ones. To avoid the computational cost of graph isomorphism solutions we use Exas [23] our efficient structural feature extraction method for graph based structure to extract a characteristic vector for each sub graph. In the algorithm as shown in the figure 3.16, each pattern P is represented by D (P), the set of its occurrences in the whole graph dataset. Each of such occurrences X is a sub graph and it might be extended into a larger sub graph by adding a new node Y and all edges connecting Y and the nodes of X. Let us denote that graph X+Y. Since a large pattern must contain a smaller pattern Y must be a frequent sub graph, i.e. an occurrence of a pattern U of size 1. This will help to avoid generating non-pattern sub graphs (i.e. cannot belong to any larger pattern).The operation $\oplus$ is used to denote the process of extending and generating all occurrences of candidate patterns from all occurrences of such two patterns P and U:

$P \oplus U = \{X + Y \mid X \in G_i(P), Y \in G_i(U), i = 1..n\}$     eq$^n$ (3.5)

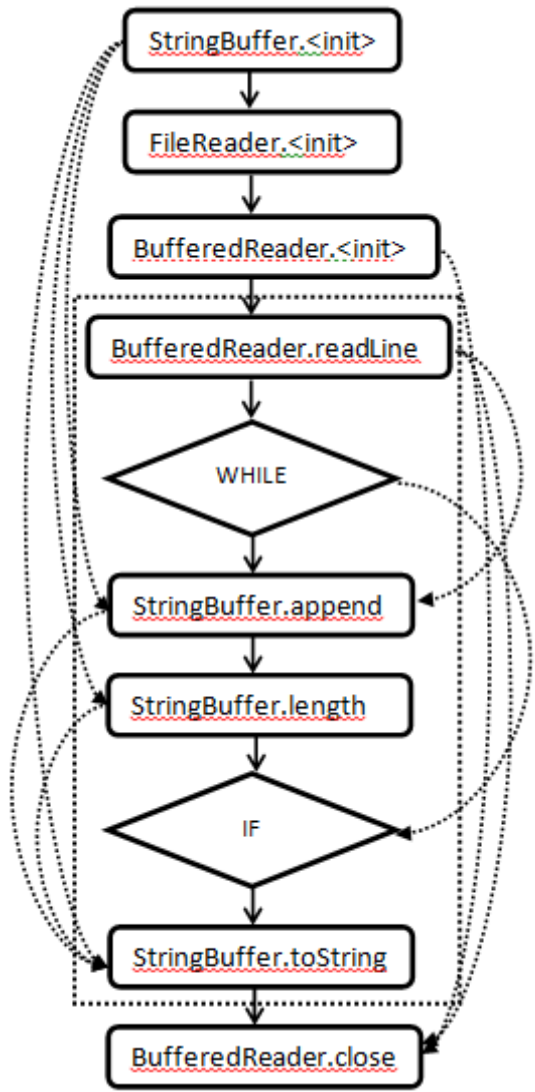

Fig.3.14: Graph based object usage model

function PatternExplorer (D)
    $L \leftarrow \{$ all patterns of size one $\}$
    for each $P \in L$ do Explore $(P, L, D)$
    return L

function Explorer $(P, L, D)$
    for each pattern of size one $U \in L$ do
        $C \leftarrow P \oplus U$
        for each $Q \in$ patterns $(C)$
            if $f(Q) \geq \sigma$ then
                $L \leftarrow L \cup \{Q\}$
                Explore $(Q, L, D)$

Fig.3.16: Pattern Explorer Algorithm (PattExplorer)

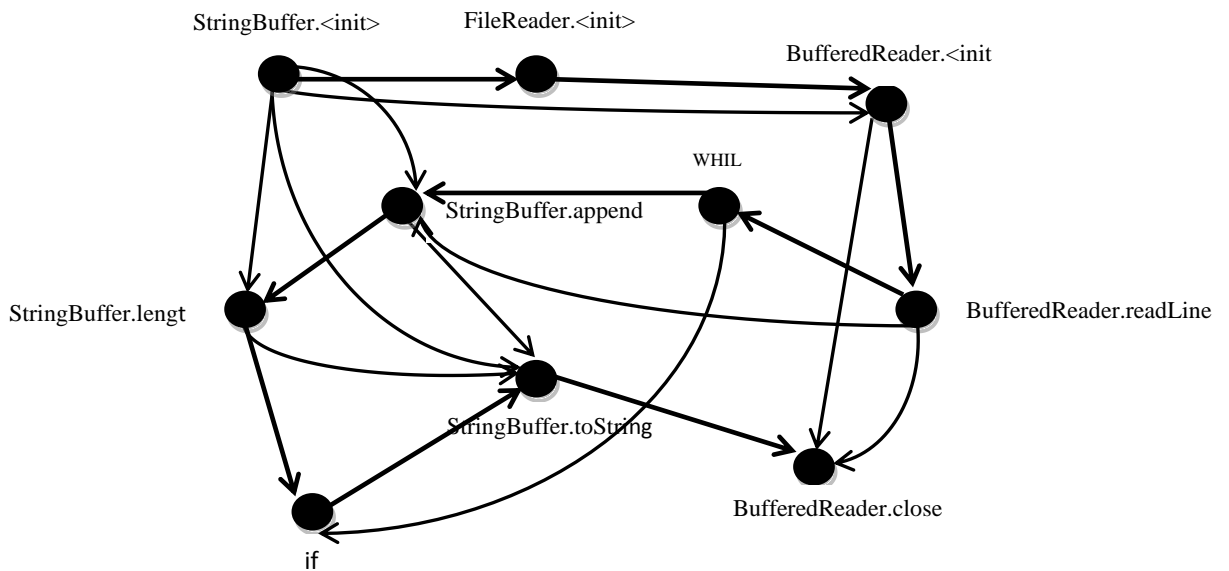

Fig.3.15 Directed Acyclic Graph of Object Usage Model

Chapter 4

# Experimental Evaluation

Enhancing Software Development Process (ESDP) aims at accelerating development process through better performance and influencing the time consumption in a platform where the integrated environment preserves the mined xml repository with better abstraction technique in server independent surroundings. To compare and justify the efficiency of ESDP system we have evaluated its response time, performance, quantitative comparison for the first and second match and empirical study. We have designed and installed our own environment before starting up the expected evaluation.

## 4.1 ENVIRONMENTAL SET UP

To evaluate performance and effectiveness of our ESDP tool, we have applied it to several open source Java projects as shown in the Table 4.1. The experiments were carried out in a computer with Windows 7 Operating System, Intel Core I 5 Processor, RAM of 4GB with 3G internet connection of local operator.

Table 4.1 Open Source Projects found from different Enterprise Repository online

| ID | Projects | Files | Methods | USG Pattern | Prominent API |
|---|---|---|---|---|---|
| P1 | *jEdit 3.0* | 14 | 74 | 8 | jmlspecsorg.eclipse.core.runtime.Plugin |
| P2 | *Log4J 1.2.15* | 17 | 79 | 4 | org.apache.commons.codec.binary.Base64 |
| P3 | *Jigsaw 2.0.5* | 11 | 28 | 7 | Javax.sql |
| P4 | *Struts 1.2.6* | 13 | 109 | 8 | oracle.core.lmx |
| P5 | *Fluid VC12.05* | 15 | 106 | 6 | com.mysql.jdbc |
| P6 | *JabRef-2.7.2* | 8 | 92 | 9 | Com.mysql.jdbc.jdbc |
| P7 | *tftp4java-0.8* | 9 | 29 | 9 | org.osgi.framework.BundleContext |
| P8 | *pooka* | 10 | 42 | 14 | com.sun.mail.auth |
| P9 | *Vocabulary Test* | 12 | 32 | 17 | javax.swing |
| P10 | *jMusic* | 6 | 32 | 7 | javax.sound.sampled |

Here, the table shows the number of files, methods, usage patterns and the Prominent API used in that project. The number files says that total number files are found in the projects, and number of methods denotes that total amount methods considered from the project sources and the similarly the usage pattern shows the total number usage pattern that we have considered that they have sequence among themselves.

## 4. 2 RESPONSE TIME EVALUATION

We have applied the projects to build our Mined API XML Repository. and Then we have searched some user queries onESDP , MAC and MAPO to check the time complexity. Here the table 6 and figure 17 show the comparisons among those found required time complexity to respond. Here, we have searched different queries on three different systems and have come to see that in case of our ESDP it takes very fewer times to get the recommendations.

Table 4.2: Response TimeComparison among different Systems

| UQ | Search Terms | MAC[9] | MAPO[5] | ESDP |
|---|---|---|---|---|
| 1 | Connection | 0.59 s | 5.39 s | 0.15 s |
| 2 | XMLParser | 0.83 s | 6.23 s | 0.10 s |
| 3 | getConnection() | 0.55 s | 5.56 s | 0.09 s |
| 4 | ActionListener | 0.74 s | 4.96 s | 0.10 s |
| 5 | InputMissmatchException | 0.67 s | 5.35 s | 0.17 s |

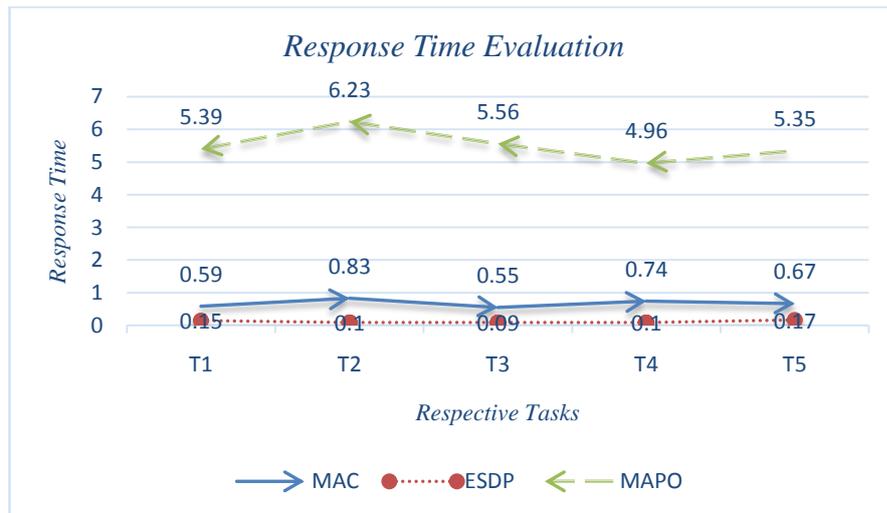

Fig. 4.1 Response Time Comparison of ESDP, MAC and MAPO

The time complexity of the ESDP tool shows that when MAC [9] takes 0.59 seconds then our ESDP takes only 0.15 seconds to respond. ESDP has amazing outcomes in case of time because of its server dependency and XML Based mining. As we know, when we want to search something online then we have to resolve the request and response latency. If the server takes longer times than usual then the developer of the system may have to wait or pass idol time that is totally inconvenient to the users. However, server dependency has another short-comes also. That is if the server fails then total system may crash and lose everything that is already developed.

## 4.3 PRECISION-RECALL PERFORMANCE EVALUATION

We have searched the user queries after building the Mined API XML repository. We also have measured how many results are indexed as search result and how many of them are relevant. As we cannot directly search the user queries on MAPO and MAC so, we cannot show their precision recall performance evaluation. We have checked our user queries on KODERS.COM and on our ESDP tool. The results are compared in the precision-recall comparative analysis.

The performance of information retrieval systems is often evaluated by analyzing their recall and precision. Precision is defined as the number of relevant materials retrieved by a search divided by the total number of materials retrieved by that search. Recall is defined as the number of relevant materials retrieved by a search divided by the total number of existing relevant materials which should have been retrieved. In our contexts, precision and recall are defined in terms of a set of retrieved API usage scenarios (e.g., the list of API usage scenarios recommended by ESDP for a query) and a set of relevant API usage scenarios (e.g., the list of API usages scenarios found in local repository by manual inspection for a particular programming task ). More formally, we define precision and recall as

$$precision = \frac{|\{relevant\ API - attribute\ usage\ pattern\} \cap \{retrieved\ API - attribute\ usage\ Pattern\}|}{|\{retrieved\ API - Attribute\ usage\ pattern\ statment\}|}$$

$$recall = \frac{|\{relevant\ API - attribute\ usage\ pattern\} \cap \{retrieved\ API - attribute\ usage\ Pattern\}|}{|\{relevant\ API - Attribute\ usage\ pattern\ statment\}|}$$

So, Precision is defined as the number of relevant materials retrieved by a search divided by the total number of materials retrieved by that search. Recall is defined as the number of relevant materials (API Sequence) retrieved by a search divided by the total number of existing relevant materials (API Sequence) which should have been retrieved. In our contexts, precision and recall are defined in terms of a set of retrieved API usage scenarios (e.g., the list of API usage scenarios recommended by ESDP for a query) and a set of relevant API usage scenarios (e.g., the list of API usages scenarios found in local repository by manual inspection for a particular programming task ). For example, more formally, we define precision and recall as below. Consider that we have a system to recognize dogs, if the system identifies 7 dogs in a scene containing 9 dogs and some cats. If 4 of the identifications are correct but 3 are actually cats then Precision is 4/7 and Recall is 4/9. When ESDP engine returns 30 API Sequences and out of total 30 API Sequences only 20 API Sequences are relevant, then the precision is P = 20/30. And as it fails to match 40 additional API Sequences that means it had 60 API Sequences items in total then Recall is, R= 20/60.Here, If we could use the association pattern rule and then ESDP recommended 5 relevant API-statements out of the top 10 recommendations where we have 6 API-Statement in total, the precision is 0.6 (6/10) and the recall is 0.83 (5/6).But as we have applied the Sequential Pattern Mining Algorithm Prefix-Span so we cannot evaluate the Precision and recall like above.  To evaluate the sequential pattern rule, we took a query statement as the input. In ESDP, each recommendation (sequence) includes several API-statements; in a query, ESDP recommended several sequences. Therefore, we compute precision and recall of each sequence, then average them in the top N. In this sample query, the relevant API-statements should contain those 4 API statements and must appear in sequence. If the ESDP recommended for the query in the first sequence, there are 5 API-statements, including four that are relevant. The precision of the sequence is 0.8 (4/5) and the recall of the sequence is 1 (4/4). But we tried to find the similar type result to extract to see the comparison.  And it has some limitations. One thing we should keep in mind that what exactly the relevance means? We have got recommended API Statement after applying a user query. And we compared that with our original API-Query API statement found checked the relevance. And the high Precision means it has more relevance than irrelevance. That anyway ensures the exactness and the Quality of the system. On the other contrary, the high Recall means that the system has the performance with most of the relevance. The high recall ensures the completeness and quantity of the system.

Table 4.3 Precision Recall for ESDP

| UQ | 5 | | 10 | | 15 | | 20 | | 25 | | 30 | |
|---|---|---|---|---|---|---|---|---|---|---|---|---|
| % | P | R | P | R | P | R | P | R | P | R | P | R |
| 1 | 100 | 20 | 80 | 32 | 85 | 52 | 75 | 60 | 72 | 64 | 70 | 84 |
| 2 | 80 | 16 | 70 | 28 | 80 | 48 | 80 | 64 | 72 | 64 | 67 | 80 |
| 3 | 60 | 16 | 70 | 32 | 80 | 48 | 70 | 56 | 64 | 57 | 57 | 68 |
| 4 | 100 | 12 | 100 | 28 | 73 | 44 | 75 | 60 | 64 | 57 | 63 | 76 |
| 5 | 40 | 20 | 60 | 40 | 93 | 56 | 75 | 60 | 64 | 57 | 53 | 64 |
| 6 | 100 | 08 | 90 | 24 | 66 | 40 | 70 | 56 | 60 | 53 | 57 | 68 |
| 7 | 80 | 20 | 80 | 36 | 86 | 52 | 75 | 60 | 76 | 79 | 67 | 80 |
| 8 | 80 | 16 | 80 | 32 | 73 | 44 | 75 | 60 | 76 | 67 | 63 | 76 |
| 9 | 100 | 16 | 90 | 32 | 80 | 48 | 70 | 56 | 68 | 60 | 67 | 80 |
| 10 | 100 | 20 | 80 | 36 | 86 | 52 | 75 | 60 | 76 | 67 | 63 | 76 |
| Top | 100 | 20 | 100 | 40 | 93 | 56 | 80 | 64 | 76 | 79 | 67 | 84 |

Table 4.4 Precision Recall for KODERS

| UQ | 5 | | 10 | | 15 | | 20 | | 25 | | 30 | |
|---|---|---|---|---|---|---|---|---|---|---|---|---|
| % | P | R | P | R | P | R | P | R | P | R | P | R |
| 1 | 80 | 16 | 70 | 28 | 73 | 48 | 70 | 56 | 64 | 57 | 57 | 76 |
| 2 | 60 | 12 | 60 | 28 | 66 | 44 | 75 | 60 | 64 | 64 | 53 | 68 |
| 3 | 60 | 12 | 60 | 28 | 80 | 48 | 70 | 56 | 56 | 53 | 53 | 68 |
| 4 | 80 | 12 | 80 | 24 | 73 | 44 | 70 | 56 | 56 | 53 | 63 | 64 |
| 5 | 60 | 16 | 60 | 36 | 80 | 48 | 75 | 56 | 64 | 57 | 53 | 64 |
| 6 | 80 | 08 | 80 | 24 | 66 | 40 | 70 | 56 | 56 | 53 | 53 | 68 |
| 7 | 60 | 16 | 70 | 32 | 66 | 48 | 70 | 60 | 64 | 57 | 63 | 76 |
| 8 | 60 | 12 | 60 | 28 | 73 | 44 | 75 | 60 | 72 | 53 | 53 | 76 |
| 9 | 80 | 12 | 80 | 28 | 80 | 48 | 70 | 56 | 56 | 60 | 57 | 76 |
| 10 | 80 | 16 | 70 | 24 | 73 | 48 | 70 | 56 | 64 | 53 | 53 | 64 |
| Top | 80 | 16 | 80 | 36 | 80 | 48 | 75 | 64 | 64 | 64 | 63 | 76 |

Table4.5: The average precision and recall for the of different user queries.

| Serial | Top Recommended Scenarios | ESDP | | KODERS | |
|---|---|---|---|---|---|
| | | Average Precision | Average Recall | Average Precision | Average Recall |
| 1 | 5 | 84.00 | 16.40 | 70.9 | 13.45 |
| 2 | 10 | 80.00 | 320 | 70.00 | 28.72 |
| 3 | 15 | 80.20 | 48.40 | 70.45 | 21.08 |
| 4 | 20 | 74.00 | 59.20 | 71.81 | 57.81 |
| 5 | 25 | 69.20 | 62.50 | 61.81 | 56.72 |
| 6 | 30 | 62.70 | 75.20 | 56.45 | 70.54 |

We present the calculated precision and recall value for each task in Table 4.3. In the Table 4.4, 'P' indicates precision and 'R' indicates recall. The comparative precision recall result is plotted in the figure 4.2 below. Here there are scenarios starting from 5 and ended to 30. The plotted curve shows that ESDP brings better relevance than the KODERS.COM.

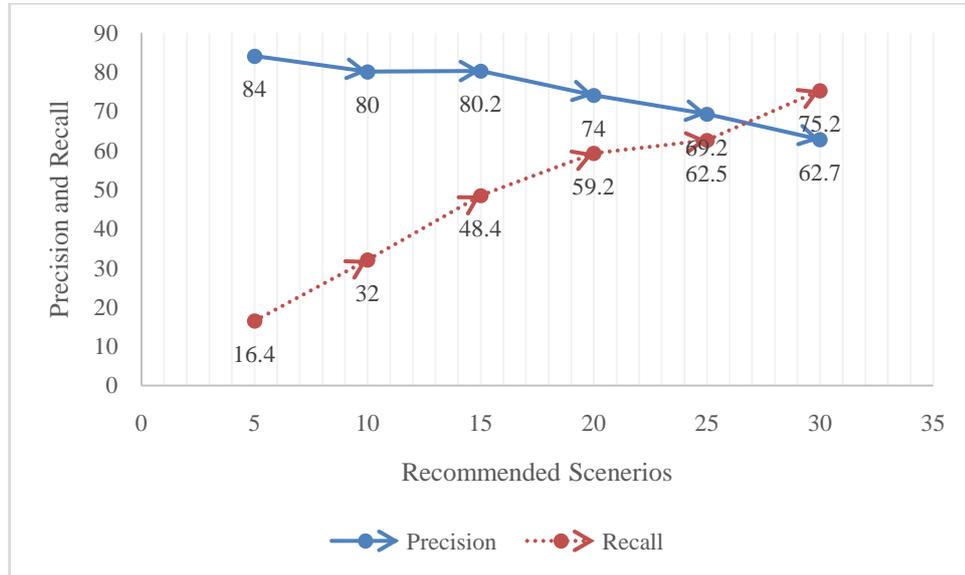

Fig. 4.2 Average Precision and Recall Plotting for ESDP

### 4.3.1 RESULT DISCUSSION OF PERFORMANCE EVALUATION

The results of the experimental evaluation to measure the performance of ESDP system we have summarized in the Table 4.3 and 4.4. For each top-N in our sample, we list the precision and recall of recommended API-statements.

And the following graph of curve shows the precision and recall values that resulted from applying our evaluation method to both investigations on ESDP and KODERS.COM. The lines connecting the data points on the recall versus precision plot show the trade-off between precision and recall as the parameter values are altered. For example, scenario 5 stands for 5 total recommendations where we have the first sequence of all the recommendations for a particular user query. Then the average is of the precision and recall for 5 to 30 scenarios is plotted in the Figure 4.2. The figure 4.2 shows that the precision decreases according to the liner increment of the scenarios whereas the recall increases if the number of scenarios is raised. And in a single point near 27 or 28 scenarios the precision and recall gets equal. The point we can call the equilibrium of precision and recall. We have considered 27 scenarios as our standard recommendations.

Each preprocess and mining task we have performed for a specific input query in this study took less time on a PC with 4 GB RAM with a 2.86 GHz processor. The task included the time it took for the ESDP to retrieve the 20 relevant source files from its Mined XML repository. On the other hand Koders.com to retrieve the 20 relevant source files from its web based database repository. For coding a program statement, the process of retrieving the relevant files is more time consuming, but it is performed less often because the MACs can find several program statements that can be used within a class or method.

As we know the Apriori algorithm computes all patterns beforehand, then searches the pattern set for a given developing task. However, computing all patterns takes time. In our experiments, ESDP uses two optimizations to compute patterns on demand. First, considering our specific application, the antecedent of the patterns is equal to the given program statement; hence, we only mine patterns which match the situation. Second, in our application, ESDP only computes patterns with a single item in their consequent. ESDP uses the PrefixSpan algorithm and dynamically controls the minimum support of mining to obtain less than 50 patterns for each query. In our experiments, 50 patterns are sufficient for our application. These optimizations make our mining very efficient.

We see that the recall and precision of the first sequence are also quite high as shown in the Table 4.3. If they are 0.82 and 0.85 respectively that means the statements of the sequence cover 82% of the developer's needs and with a precision 85%. When the sequence length was reduced, the recall was reduced too, but precision rose. Shorter sequences cannot cover all necessary statements, but they can do it more easily target the need.

Either Association or sequential patterns mining both generally assume a large number of transactions. Too few transactions may affect the results of mining. To address the performance issue, we retrieve 20 relevant source files and produce transaction in the mined XML repository for further sequential pattern mining. The number of transactions seems too few; however, our objective in mining is to find out the limit API-usage patterns that are relevant to the statement assigned by the developer, rather than all API-usage patterns. In fact, in our study, if we have retrieved too many relevant source files, some useful patterns related to the query may be lose due to the minimum support count set to achieve better performance. So we can say throughout the precision and recall evaluation that the Enhancing Software Development Process (ESDP) works much better than many of the related and existing approach in the field of mining software repository. We also like to mention that the data available here are collected from manual inspection using the ESDP and other system and their repository.

**4.3.2 RECEIVER OPERATING CHARACTERISTICS (ROC) CURVE ANALYSIS**

Receiver Operating Characteristic (ROC), or simply ROC curve, is a graphical plot which illustrates the performance of a binary classifier system as its discrimination threshold is varied. It is created by plotting the fraction of true positives out of the positives (TPR = true positive rate) vs. the fraction of false positives out of the negatives (FPR = false positive rate), at various threshold settings. True Positive (TP) is the correct matches as we want to. False Negative (FN) is the non-matches item that is correctly rejected.

$$TPR = \frac{TP}{TP+FN} \text{ (Eqn. 4.1)}$$

$$FPR = \frac{FP}{FP+TN} \text{ (Eqn. 4.2)}$$

Therefore, $Precision = \frac{TP}{TP+FP}$ (Eqn. 4.3) and

$$Recall = \frac{TP}{TP+FN} \text{ (Eqn. 4.4)}$$

TPR is also known as sensitivity, and FPR is one minus the specificity or true negative rate. ROC analysis is related in a direct and natural way to cost/benefit analysis of diagnostic decision making [27].

In our performance evaluation we have calculated the precision and recall in different cases. Then we have drawn the ROC curve. The datasets are built after executing different queries by turns.

The ROC is drawn and shown in the figure 4.3. The blue dotted line marks ROC for ESDP system and red dashed line marks the ROC curve for the KODERS. From the graph we see that the standard region for the ESDP is better than the standard region of KODERS.

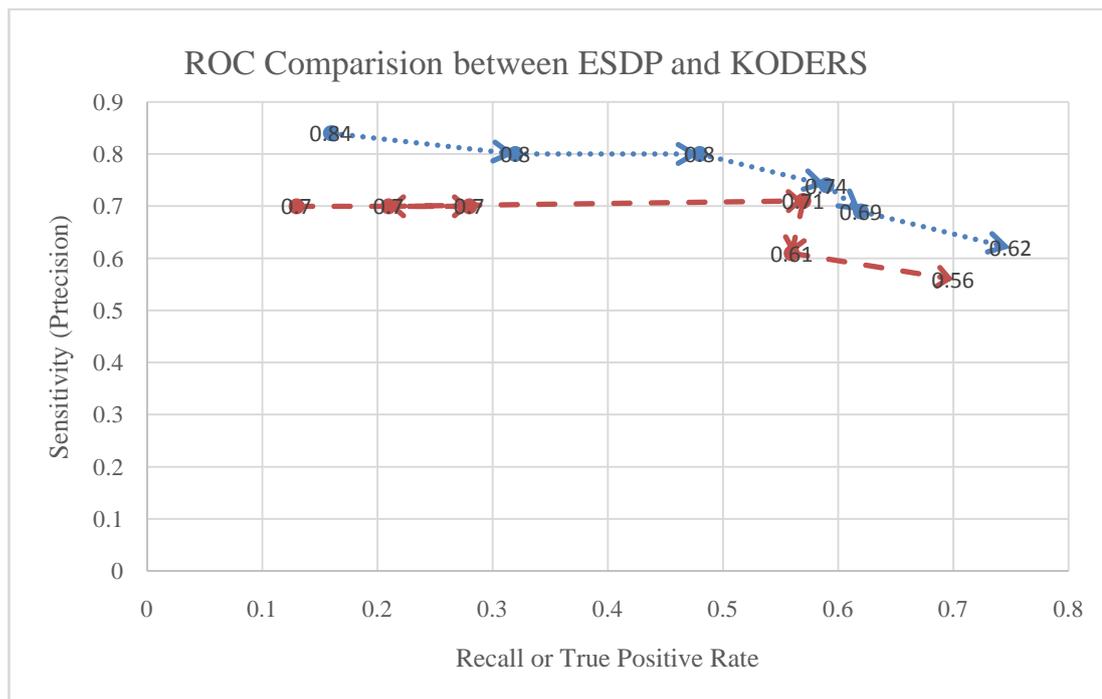

Figure 4.3: ROC Comparison between ESDP and KODERS

## 4.4 QUANTITATIVE COMPARISON

We have taken five different user queries (UQ) to see the search matching. When we search for the first user query the result for the first match is found within 6 results by MAC [9], within 2 results by MAPO[4][5] and within 1 result by our ESDP tool. The entire system is manually designed for experiment purpose. Similarly for the other user queries the found results for both first matching and second matching is listed in the following table 4.9.

Table 4.10 Comparative results of the three tools to locate the 1st and 2nd matches

| UQ | First Matched | | | Second Matched | | | Sum of the items | | |
|---|---|---|---|---|---|---|---|---|---|
|  | MAC | MAPO | ESDP | MAC | MAPO | ESDP | MAC | MAPO | ESDP |
| UQ 1 | 6 | 2 | 1 | 7 | 3 | 2 | 13 | 5 | 3 |
| UQ 2 | 2 | 1 | 1 | 3 | - | 1 | 5 | 1 | 2 |
| UQ 3 | 2 | 2 | 2 | 2 | 3 | 2 | 4 | 5 | 4 |
| UQ 4 | - | 1 | 1 | 1 | 2 | 1 | 1 | 3 | 2 |
| UQ 5 | 2 | - | - | - | 2 | 1 | 2 | 2 | 1 |

## 4.5 EMPIRICAL STUDY

To see the efficiency with respect to the error vulnerability we have formed two different teams. The first team is called Experimental Team and second team is called ESDP team. Both teams have two members. Then they are said to do three tasks. The experimental team use the manual process using Intellij Idea[25] IDE and the second team use the IDE where ESDP tool is integrated. The three different tasks are given in the following table 4.10. For the first experiment the Experimental Team consisting of members 1 and 2 brings 6 errors while they do their given tasks where as the ESDP teams face only two according to our observation as shown in the table 4.11 and 4.12.

Table 4.11 Three tasks given to two teams to find out the error vulnerability

| Task ID | Description | API Calls |
|---------|-------------|-----------|
| T-001 | Add a context menu to an editor | 5 |
| T-002 | Update the name and the bounds of a figure | 4 |
| T-003 | Save the content of a editor | 8 |

Table 4.12 Result of the Empirical Experiment I

| Tasks | Experimental Team | | | ESDP Team | | |
|-------|----------|----------|-------|----------|----------|-------|
|  | Member 1 | Member 2 | Total | Member 3 | Member 4 | Total |
| T-001 | 0 | 1 | 1 | 1 | 0 | 1 |
| T-002 | 1 | 2 | 3 | 0 | 1 | 1 |
| T-003 | 0 | 2 | 2 | 0 | 0 | 0 |
| Grand Total= | | | 6 | Grand Total= | | 2 |

For the second experiment team members are shuffled. That means the Member 1 and two works for the ESDP team and the rest of the Member 3 and 4 works for the Experiment team. And in that case the given tasks were not changed that they are said to do the same tasks.

Table 4.13 Result of the Empirical Experiment II

| Tasks | ESDP Team | | | Experimental Team | | |
|-------|----------|----------|-------|----------|----------|-------|
|  | Member 1 | Member 2 | Total | Member 3 | Member 4 | Total |
| T-001 | 2 | 2 | 4 | 3 | 2 | 5 |
| T-002 | 1 | 3 | 4 | 5 | 3 | 8 |
| T-003 | 0 | 2 | 2 | 4 | 3 | 7 |
| Grand Total= | | | 10 | Grand Total= | | 20 |

From the Empirical Study we see that the ESDP teams face less number of errors than the Experimental teams. So we can say that our system works better that the usual process. This is shown in the table 16.

# Chapter 5

# Conclusion and Future Work

The application of data mining technique is presented as a task that can offer great advantages and potentials. There are number of efforts have been seen to speed up software coding process within the least amount of time. This thesis titled "Enhancing Software Development Process (ESDP) using Data Mining Integrated Environment" is actually a method for the application of data mining tools that work with the general data mining technique with graph mining algorithm to guide developers in the coding phase of software development. As we claim ourselves that our ESDP system has better source code abstraction with a smooth representation using XML parsing process in mining and server independency in searching beside enriched repository than any other existing approaches. Through the work we have ever done the ESDP framework is able to recommend the snippets more accurately. For data mining we have used prior mining process instead of post mining [9]. That means our source abstraction is mined as a whole using data mining algorithm and then it is stored in the XML data storage. For this reason it is able to work in the client system without being interfered by any server dependency. In response to the user query it provides recommendation sequences of usage pattern with necessary code skeleton in background. Then the code skeleton is refactored and updated with the snippets it needed. After that a graph based algorithm pattern explorer and EXUS is used to provide the final recommendation to the user.

We also investigated how implicit Class frame, Methods and API-usage patterns mined from relevant source files can be used to facilitate software development. To investigate our approach with realistic programming tasks, we have designed an ESDP Tool that works as a data mining integrated environment. Using ESDP Tools can form such reference Frames for the Classes, Methods and API-usage pattern along with the recommended patterns suggested from the databases.

In this thesis we have tried to enhance the software development process through reusing the usage pattern of existing projects. It is a basic need to reuse existing Application Programming Interface (API), Class Libraries or even framework for rapid software development. Our ESDP system is able to ease this process through the framework.

ESPD concept is different from most other approaches mentioned above in many extents. In future we intend to work with our ESDP system purposing professional implementation after eliminating the flaws that are found in our current research.

## 5.1 FUTURE WORK

In future the research can be implemented for the use of professional purposes. Here, we have evaluated the efficiency by collecting codes of some open source projects, but before we implement this concept professionally the empirical evaluation should be done by using large scale projects. ESDP tool is developed considering the research purpose use. To extend the use it needs to be tested professional testing tool.

Another thing is that there is a rich literature regarding bug detection and prediction. Most studies used the analysis of static code to detect bugs in software systems. In Recent years, with the increasing usage of data mining, mining bugs from source control systems has become one of the more rapidly advancing subfields of mining software repositories. In future, we will try to mine existing projects that will be used to detect the bugs and errors in the proposed applications. Thus it will improve user experience.

# Appendix

## APPENDIX A

To evaluate performance and effectiveness of our ESDP tool, set up our system with the following configuration-

- System: Intel Core I 5 Processor
- Operating System: Windows 7 Operating System
- RAM: 4 GB
- Tested Projects: Open Source Java projects as shown in the Table 5

## APPENDIX B

As we are using PattExplorer (Pattern Explorer) algorithm we have to use some general terms related to this algorithm.

*DEFINITION 1* (Groum): *A groum (Graph-Based Object Usage Model) is a DAG such that:*

- *Each node is an action node or a control node.*
- *A groum could involve multiple objects.*
- *Each edge represents a (temporal) usage order and a data dependency*

*DEFINITION 2: Two groums are (semantically) equivalent if they are label-isomorphic [MOUP 24].*

*DEFINITION 3: A groum dataset is a set of all groums extracted from the code base, denoted by $D = \{G_1, G_2, ...,G_n\}$.*

*DEFINITION 4: An induced sub graph X of a groum $G_i$ is called an occurrence of a groum P if X is equivalent to P.*

***DEFINITION 5:****The frequency of P in $G_i$, denoted by $f_i(P)$, is the maximum number of independent (i.e. non-overlapping) occurrences of P in $G_i$. The frequency of P in the entire dataset, f(P), is the sum of frequencies of P in all groums in the dataset.*

***DEFINITION 6 (PATTERN):*** *A groum P is called a pattern if $f(P) \geq \sigma$, i.e. P has independently occurred at least σ times in the entire groum dataset. σ is a chosen threshold.*

***DEFINITION 7 (PATTERN MINING PROBLEM):****Given D and σ, find the list L of all patterns.*